\documentclass[aps, prl, twocolumn, superscriptaddress]{revtex4-1} 
\usepackage{amssymb, amsmath, amsthm}
\usepackage{xcolor}
\usepackage{graphicx}
\usepackage{hyperref}
\usepackage{comment}
\usepackage{blkarray}
\usepackage{mathtools}
\usepackage{tikz}
\usetikzlibrary{decorations.pathreplacing}
\usepackage{dsfont}
\usepackage{braket}

\usepackage{graphicx}
\usepackage{dcolumn}
\usepackage{bm}

\newcommand{\prlsection}[1]{{\em {#1}.---~}}
\newcommand{\Tr}[1]{\mathrm{Tr}}

\newcommand{\ii}{{\rm i}}
\newcommand{\dd}{{\rm d}}

\newcommand{\titleinfo}{
Squeezed ensembles and anomalous dynamic roughening\\
in interacting integrable chains}
\begin{document}

\preprint{APS/123-QED}

\title{\titleinfo
}

\author{Guillaume Cecile}
\affiliation{Laboratoire de Physique Th\'eorique et Mod\'elisation, CNRS UMR 8089,
	CY Cergy Paris Universit\'e, 95302 Cergy-Pontoise Cedex, France}

\author{Jacopo De Nardis}
\affiliation{Laboratoire de Physique Th\'eorique et Mod\'elisation, CNRS UMR 8089,
	CY Cergy Paris Universit\'e, 95302 Cergy-Pontoise Cedex, France}

\author{Enej Ilievski}
\affiliation{Faculty of Mathematics and Physics, University of Ljubljana, Jadranska 19, 1000 Ljubljana, Slovenia}

\date{\today}

\begin{abstract}
	It is widely accepted that local subsystems in isolated integrable quantum systems equilibrate to generalized Gibbs ensembles. Here, we demonstrate the failure of canonical generalized thermalization for a particular class of initial states in certain types of interacting integrable models. Particularly, we show that in the easy-axis regime of the quantum XXZ chain, pure non-equilibrium initial states with no magnetic fluctuations instead locally relax to squeezed generalized Gibbs ensembles, referring to exotic equilibrium states governed by non-local equilibrium Hamiltonians with sub-extensive charge fluctuations that violate the self-affine scaling. The behaviour at the isotropic point is exceptional and depends on the initial state. We find that relaxation from the N\'{e}el state is governed by extensive fluctuations and a super-diffusive dynamical exponent compatible with the Kardar-Parisi-Zhang universality. On the other hand, there are other non-fluctuating initial states that display diffusive scaling. Our predictions can be directly tested in state-of-the-art cold atomic experimental settings.
\end{abstract}

\maketitle

\prlsection{Introduction}
The study of nonequilibrium dynamical properties in isolated quantum many-body systems resulting from pure states has been at the forefront of theoretical and experimental research in the past decade \cite{polkovnikov2011colloquium,deutsch1991,srednicki1994chaos,calabrese2006time,calabrese2007quantum,rigol2008thermalization,cazalilla2010focus,fagotti2013reduced,calabrese2012quantum,Kinoshita2006,caux2016quench,caux2013time,essler2016quench,dalessio2016quantum,cotler2021emergentquantum,piroli2016mutiparticle,PhysRevX.12.041032,ho2022exact,trotzky2012probing,Langen_2015,Lange2017,PhysRevLett.129.106802,Wang_2022,Malvania_2021,Morvan2022}. Quantum quench protocols have provided a versatile and fruitful tool for understanding the key mechanisms leading to thermalization: while the global state remains pure at all times, the reduced density matrix of any large sub-system typically evolves at late times to a maximally entropic state subject to the constraints of local conserved quantities. This paradigm has been examined in a rich variety of systems, including generic chaotic models and free or interacting integrable models. Concurrently, there have been important developments in understanding the eigenstate thermalization hypothesis \cite{cazalilla2010focus,srednicki1994chaos,dalessio2016quantum,PhysRevLett.129.170603} and its generalization to the integrable cases \cite{lydzba2021entanglement,Wang_2022,Buca23}.  
\begin{figure}
	\includegraphics[scale=0.39]{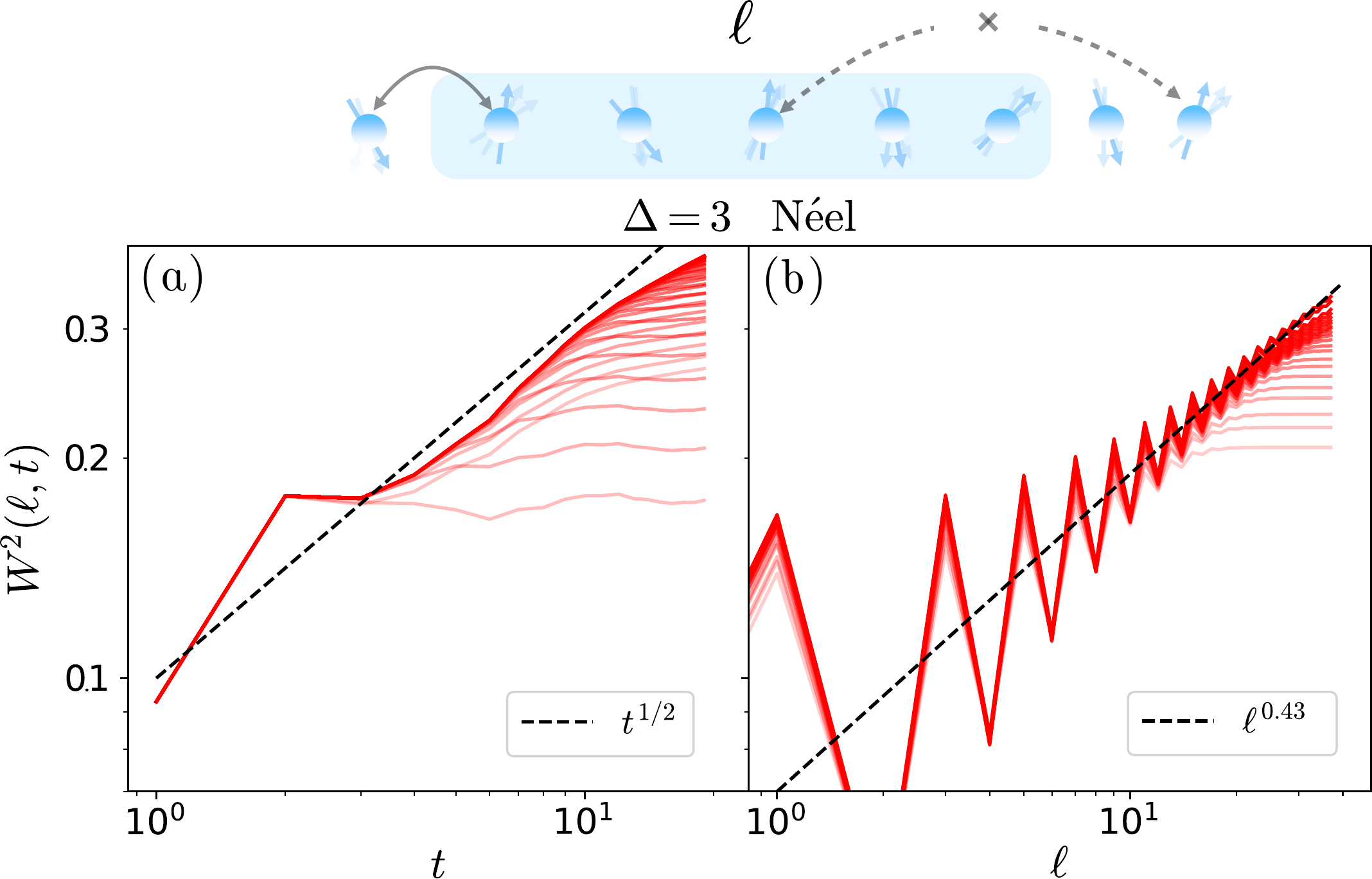}
	\caption{Top: non-equilibrium time evolution from an initial state. Spatial fluctuations of local charge $Q_\ell$ produced during the time evolution
    in the subsystem of length $\ell$ grow sub-extensively with for large $\ell$. 
    Bottom: temporal and spatial scaling of magnetic fluctuations after the quench from the Néel state in the Heisenberg XXZ chain with $\Delta=3$: (a) double log-plot of $W^{2}(\ell,t)$ as function of time for different $\ell \in [4,40]$ (increasing from light to dark); (b) double log-plot of $W^{2}(\ell,t)$ as function of $\ell$ for different times $t \in [5,16]$ with $\delta t=0.5$ (from light to dark), with the asymptotic scaling $W^{2}(\ell,t) \sim \ell^{2\zeta}$ and fitted exponent $2\zeta \approx 0.43$.  Analogous results for the Dimer initial state are reported in \cite{SM}.}
	\label{fig:sketch}
\end{figure}

Thermalization in isolated extended systems primarily concerns the stationary values of local observables following a quench from a pure initial state. Local, but large, subsystems thermalize whenever the reduced density matrices are described by canonical Gibbs ensembles or, in the case of integrable models, the generalized Gibbs ensembles \cite{rigol2008thermalization,vidmar2016generalized,essler2016quench}, characterized by finite densities of local charges and possess extensive, strictly positive, fluctuations (i.e. static charge susceptibilities).

In this work, we revisit the problem of thermalization in integrable models. We specifically consider interacting quantum spin chains with a global $U(1)$ charge $Q$ (e.g. magnetization in spin chains or electron charge in interacting fermions) and confine our study to quenches from non-equilibrium initial states belonging to a specific charge sector, i.e. superpositions of degenerate eigenstates of $Q$. In contrast with the widespread belief, we find that the reduced density matrix emerging at late times is \emph{not} a faithful GGE, but rather a non-canonical ensemble generated by a super-extensive effective Hamiltonian. Such ensembles possess a \emph{divergent} $U(1)$ chemical potential indicative of sub-extensive charge fluctuations. We dub such states suggestively as \emph{squeezed GGEs} (SGGEs).

To substantiate our claims, we consider the anisotropic (XXZ) Heisenberg chain,
\begin{equation}
    H = \sum_j \left[ S_j^x S_{j+1}^x + S_j^y S_{j+1}^y + \Delta S_j^z S_{j+1}^z \right],
\end{equation}
where $S^{\alpha}_j$ are the spin-$1/2$ generators and $\Delta$ is the interaction anisotropy. We shall mostly be interested in the $\Delta > 1$ regime, where we provide robust evidence that initial states without (global) magnetic fluctuations equilibrate locally to equilibrium ensembles $\rho_{\ell}$ with \emph{sub-extensive} (in the sub-system size) fluctuations, thus possessing vanishing static spin susceptibility $\chi=0$.  We argue that such ensembles cannot be captured by canonical GGEs generated by quasilocal effective equilibrium Hamiltonians. To support that, we demonstrate \cite{SM} that the rescaled (Hilbert--Schmidt) norm $\|\log \varrho_{\ell}\|/\ell$ \emph{diverges} with $\ell$ (see \cite{SM}). Such an anomalous behavior disappears in the gapless regime $\Delta<1$, where the spectrum of quasiparticles comprises only finitely many magnon species, consistently with thermalization to canonical GGEs. Curiously, at the isotropic point $\Delta=1$ we encounter a qualitatively different behaviour. While some initial states, e.g. the antiferromagnetic Néel state, are found to comply with canonical GGE description, with finite magnetic susceptibility and super-diffusive spin transport, other non-fluctuating pure states (e.g. the product state of spin singlets) that reveal very distinct, unorthodox properties.

Given a pure initial state $\ket{\Psi}$, we probe magnetic fluctuations within finite sub-lattices $\Lambda_{\ell}$ of size $\ell$, $Q_\ell = \sum_{j\in \Lambda_{\ell}} S_j^z$, and investigate the dynamical scaling properties of the local second moment
\begin{equation}\label{eq:variance}
W^{2}(\ell,t)\equiv \bra{\Psi(t)}Q^2_\ell \ket{\Psi(t)}.    
\end{equation}
Drawing an analogy with the interface roughness in stochastic models of interface growth, the general expectation is that $W(\ell,t)$ exhibits a self-affine Family-Vicsek (FV) scaling form \cite{PhysRevLett.52.1669,FamilyVicsek,Vicsek,PhysRevLett.69.929,Takeuchi2018,PhysRevLett.124.210604},
\begin{equation}\label{eq:growthell}
    W(\ell,t) \sim \ell^{\zeta} \, \varPhi(t/\ell^z),
\end{equation}
with $\varPhi(y)\sim y^{\beta}$ for $y\ll 1$ and $\varPhi(y)\to 1$ for $y\gg 1$, roughness (Hurst) exponent $\zeta$, and growth exponent $\beta=\zeta/z$. While in the case of e.g. N\'{e}el state we confirm the above scaling, both at the isotropic point, with Kardar-Parisi-Zhang (KPZ) exponents $\zeta=1/2$, $z=3/2$ \cite{PhysRevLett.56.889}, and for $\Delta<1$ (with ballistic exponents $\zeta=1/2$, $z=1$), we observe violation in the diffusive regime ($z=2$), where $W^{2}(\ell,t)$ scales sub-extensively with $\ell$, with an estimated (fitted) exponent is $\zeta\approx 0.22<1/2$ at $\Delta = 3$. While at the present we have no theory to predict the values of roughness exponents $\zeta \leq 1/2$, we have verified that they dependent on anisotropy and, possibly, also on the type of initial state, see additional plots in \cite{SM}. 

\prlsection{GGEs for interacting integrable systems}
In the scope of the standard quantum quench protocol, we consider integrable interacting quantum spin chains with an internal (charge) degree of freedom. For simplicity, we assume the system possesses a single $U(1)$ charge (i.e. no nesting) and consider only a class of product pure initial states $\ket{\Psi}$ of the form $\ket{\Psi}=\ket{\psi}^{\otimes n}$, with system length $L$ and  $n=L/b \in \mathbb{N}$, where $\ket{\psi}$ is a `block state' involving $b$ adjacent lattice sites. In the thermodynamic limit, the main object of interest is the reduced density matrix on a sublattice $\Lambda_{\ell}$ of size $\ell$,
$\varrho_\ell(t) = \lim_{L\to \infty} {\rm Tr}_{\bar{\Lambda}_{\ell}} \ket{\Psi(t)} \bra{\Psi(t)}$,
where the trace is over the complementary lattice $\bar{\Lambda}_{\ell}$. At late times, $\varrho_{\ell}(t)$ is expected to relax towards a GGE, $\lim_{\ell \to \infty}\lim_{t\to \infty}\varrho_\ell(t) = \varrho_{\rm GGE}$, involving in general all (quasi)local conserved quantities $I_{i}$ of the model \cite{PhysRevLett.115.120601,ilievski2016quasilocal} (coupling to chemical potentials $\beta_i$) and global $U(1)$ charge $Q$, namely $\varrho_{\rm GGE} = \mathcal{Z}^{-1}_{L}\exp{(-\sum_{i}\lambda_{i}I_{i}+h\,Q)}$ (cf. \cite{SM} for a precise definition).

Generalized Gibbs ensembles admit several equivalent descriptions \cite{ilievski2017from,10.21468/SciPostPhys.7.3.033}. One can for instance employ various state functions of the Thermodynamic Bethe Ansatz (TBA) enumerated by (integer) quantum numbers $s$, e.g. the macrostate densities $\rho_{s}(u)$ of quasiparticles with (bare) momenta $k_{s}(u)$ with rapidity $u$, or Fermi occupation (filling) functions $n_{s}(u) = \rho_{s}(u)/\rho^{\rm tot}_{s}(u)$, where the total density of states $\rho^{\rm tot}_{s}(u)$ are related to dressed momenta $p_{s}$ via $\rho^{\rm tot}_{s}(u) = p^{\prime}_{s}(u)/2\pi$.
Crucially, the coarse-grained information stored in state functions $\rho_{s}(u)$ is sufficient to uniquely fix all local correlation functions in a GGE \cite{pozsgay2014correlations,Mestyan2014}. 

There is an important subclass of initial states with $b=2$, playing the role of integrable reflecting boundaries of an integrable bulk theory \cite{Pozsgay2013,Pozsgay2018}. Such states are particularly convenient as they permit analytic closed-form computation of the GGE state functions. For general states $\ket{\Psi}$ with $b>2$, the task boils down to evaluating the expectation values $\bra{\Psi}I_{i}\ket{\Psi}$ and infer the quasiparticle densities with aid of the string-charge duality \cite{ilievski2016string}. The first successful demonstration of this program has been achieved in Ref.~\cite{ilievski2015complete} (see also \cite{ilievski2016string,pozsgay2017generalized,piroli2017correlations,piroli2016exact,Vernier2017,ilievski2017from}), thereby solidifying the concept of GGEs in integrable interacting theories. However, Ref.~\cite{ilievski2015complete} erroneously identifies the emergent ensemble as a GGE \footnote{The entire computational procedure is nevertheless correct, yielding the correct values of local correlators.}: as we clarify in turn, the vanishing of static spin susceptibility is a signature of a non-canonical ensemble.

To provide a few instructive examples, we subsequently focus on the Heisenberg spin chain, confining our analysis to the $\Delta \geq 1$ regime, where the excitation spectrum (above the ferromagnetic vacuum) comprises an infinite tower of magnon bound states with $s=1,2,\ldots$ quanta of magnetization.
The density of free energy $f=-\lim_{L\to \infty}L^{-1}\log \mathcal{Z}_{L}$ in \emph{canonical} GGEs can be split as
$f = h/2 - \mathfrak{f}$, with \cite{SM}
\begin{equation}\label{eq:infsum}
    \mathfrak{f} \equiv -\sum_{s=1}^{\infty} \int^{\pi/2}_{-\pi/2}
    \frac{\mathrm{d} u}{2\pi}  k^{\prime}_{s}(u) \log \Big(1-n_{s}(u) \Big).
\end{equation}
Noticing that the kernel $k^{\prime}_{s}(u)$ in the integrand tends to a constant at large $s$, the convergence of the infinite sum \eqref{eq:infsum} is fully predicated on the large-$s$ behavior of $n_{s}$.

\prlsection{Squeezed GGEs}
We consider quantum quenches from a class of pure initial states $\ket{\Psi}$ with vanishing charge cumulants, $c^{(n)} = (\mathrm{d}/\mathrm{d}
\lambda)^{n}F_{Q}(\lambda)|_{\lambda=0}=0$,
with the scaled cumulant generating function $F_{Q}(\lambda)\equiv \lim_{L\to \infty} L^{-1}\log \bra{\Psi}e^{\lambda\,Q}\ket{\Psi}$.
In spite of $c^{(n)}$ remaining globally conserved at all times, any subsystem of length $\ell$ will in general possess positive time-dependent cumulants $W^{(n)}(\ell,t)= \bra{\Psi(t)}Q^{n}_{\ell}\ket{\Psi(t)}$. Accordingly, one expects the emergent local equilibrium state to exhibit strictly positive
cumulants, namely $\chi^{(n)} = \lim_{\ell \to \infty}\lim_{t\to \infty}\ell^{-1}W^{(n)}(\ell,t)>0$. Surprisingly, however, this is not what happens in the easy axis regime $\Delta>1$, where $W^{2}(\ell,t)$ instead behaves anomalously, scaling sub-extensively with $\ell$.

\begin{figure}
	\includegraphics[scale=0.6]{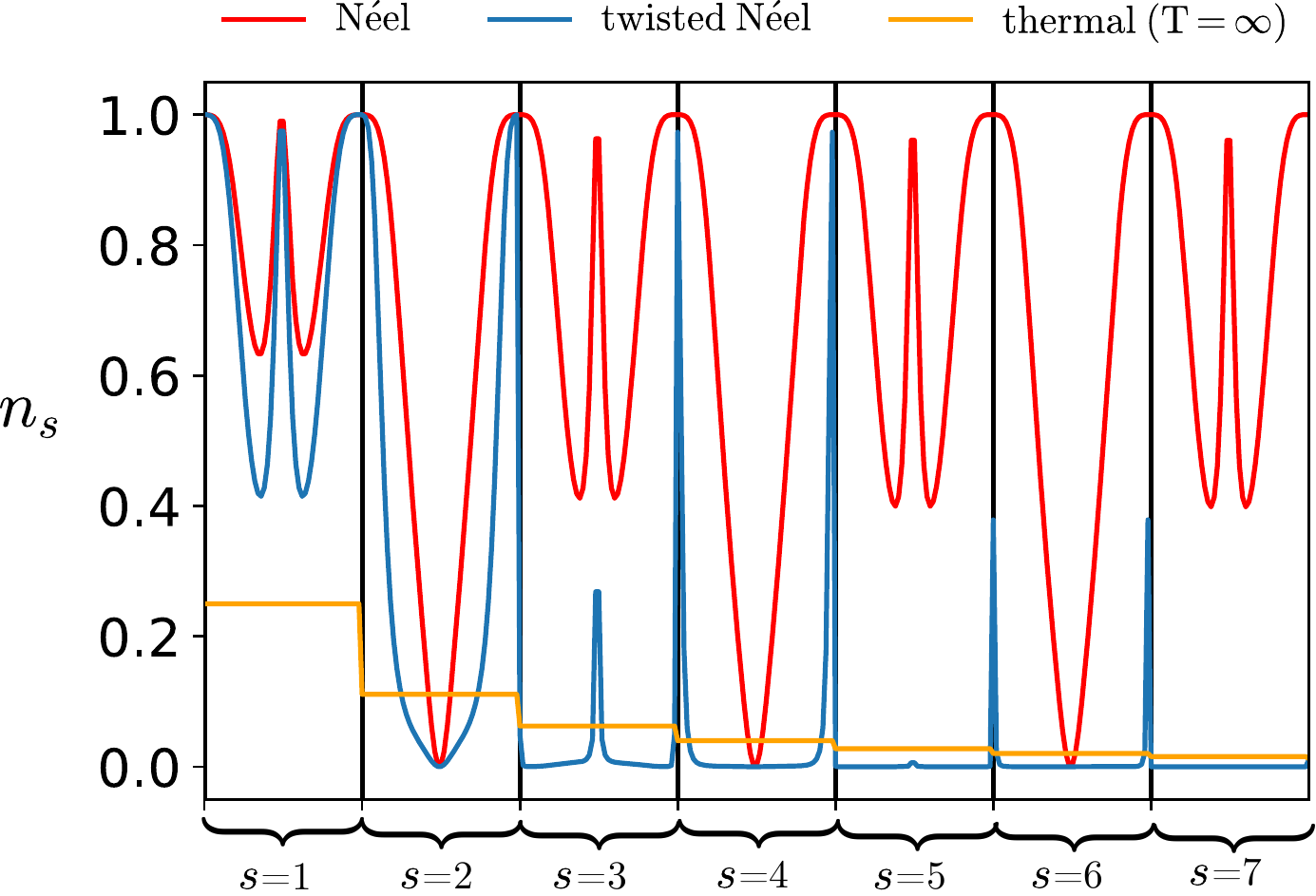}
	\caption{
 Freezing of the quasiparticle mode occupations $n_s(u)$ in the SGGE emerging from the N\'{e}el quench in the XXZ chain with $\Delta=2$, showing the Brillouin zones with $u \in [-\pi/2,\pi/2]$ (delimited by black vertical lines) for the few initial $s$, and compared to canonical behavior in GGEs (shown for the twisted N\'{e}el state $\ket{\Psi_{N}(\tau)}$) and thermal equilibrium values in the high-temperature limit.}\label{fig:ns}
\end{figure}

We now shortly discuss how such a dynamical suppression of magnetic fluctuations in local equilibrium states is subtly related to `\emph{freezing}' of the mode occupations $n_{s}$; instead of diminishing with increasing $s$, $n_{s}$ are found to converge toward non-trivial limiting functions (attractors) depending on whether $s$ is even or odd, see. Fig.~\ref{fig:ns}.
As a corollary, the infinite sum over the quasiparticle spectrum becomes divergent, $\mathfrak{f} \to \infty$. The free-energy density $f$ nonetheless remains finite. Indeed, infinitely many contributions can be resumed using certain kernel identities (cf. \cite{SM} for details), signifying that $f$ is manifestly finite in both canonical GGEs and squeezed ensembles. Importantly, however, a \emph{divergent} $\mathfrak{f}$ implies $h=\infty$. In simple terms, a singular $U(1)$ chemical potential means that the steady-state reduced density matrix is subject to a microcanonical constraint, i.e. confined within a fixed magnetisation sector of the Hilbert space.

Before heading on to explicit examples, there are several key remarks in order: (i) although the employed TBA formulae are strictly applicable only for canonical GGEs \cite{ilievski2017from,10.21468/SciPostPhys.7.3.033}, one can always regularize a divergent $\mathfrak{f}$ by introducing an appropriate twist (say $\tau$) that renders $\mathfrak{f}$ and hence also $\chi^{(n)}$ finite, removing the twist only at the end; (ii) we emphasize that $\langle Q \rangle=0$ does not generally imply $h=0$ in a GGE. In fact, for $\Delta>1$ and $\tau>0$ we find instead $\mathfrak{f}(\tau)<\infty$, but with $\mathfrak{f}(\tau)$ and $h=h(\tau)$ both diverging as $\tau \to 0$, see \cite{SM}; (iii) the peculiar freezing phenomenon cannot take place in integrable systems with a finite number of bound states since $\mathfrak{f}$ cannot grow unboundedly. Hence, there is no freezing taking place in the gapless regime with $|\Delta|<1$ and, for the same reason, this effect is genuinely due to attractive interaction (see also \cite{SM}); (iv) our conclusions apply likewise to non-fluctuating magnetized states with $c^{(1)} \neq 0$ upon subtracting the first moment in eq.~\eqref{eq:variance}, $Q_\ell \rightarrow Q_\ell - \langle Q_\ell \rangle$.

Proceeding now to explicit examples, we focus our analysis on simple initial valence-bond product states of two-site ($b=2$) blocks \cite{Pozsgay2018}. We consider specifically the N\'{e}el state and the ``Dimer" state,  
\begin{equation}
    \ket{\Psi_N} =\ket{\uparrow \downarrow}^{\otimes L/2}, \quad \ket{\Psi_D} =  \Big[ \frac{\ket{\uparrow \downarrow} - \ket{\downarrow \uparrow}}{\sqrt{2}}\Big]^{\otimes L/2},
\end{equation}
allowing for explicit analytic computation of state functions in a recursive manner \cite{SM}.
Our main conclusions nevertheless hold very generally, i.e. are valid for other initial product states that only involve eigenstates with the same value of $Q$. We have verified that the observed anomalous relaxation is not an artefact of coherent pair-production associated with integrable quenches \cite{Piroli2017,10.21468/SciPostPhys.6.5.062}.

\begin{figure*}[t]
	\includegraphics[scale=0.5]{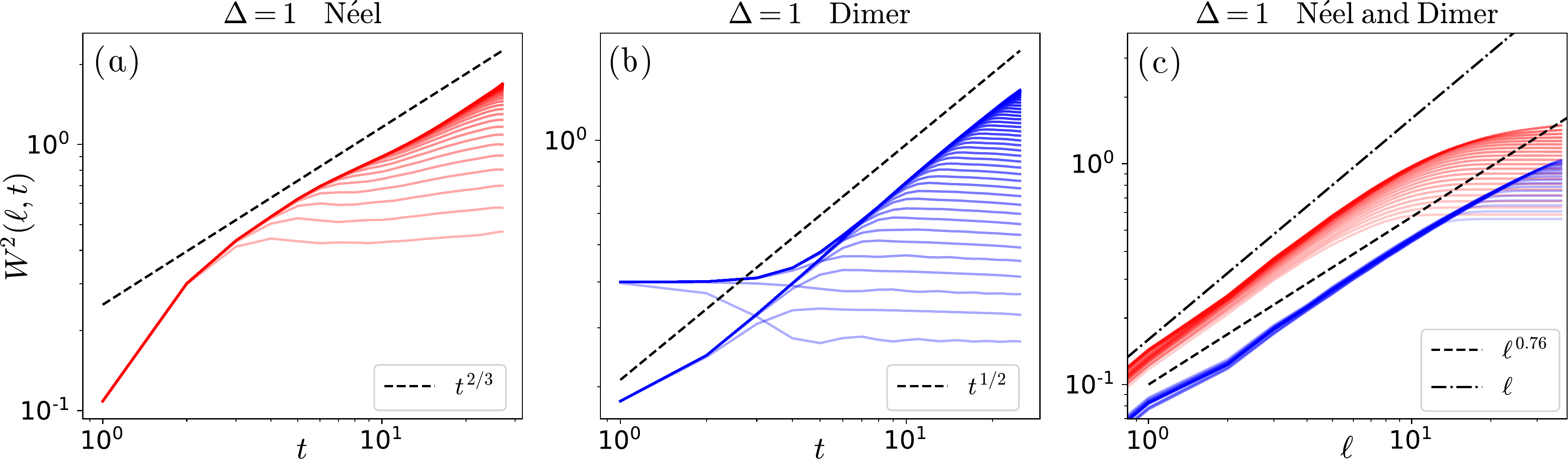}
	\caption{Local magnetic fluctuations $W^{2}(\ell,t)$ in the isotropic Heisenberg chain ($\Delta =1$): double log-plot of $W^{2}(\ell,t)$ as a function of $t$ for $\ell \in [2,29]$ for (a) the N\'{e}el and (b) Dimer initial states, compared to $t^{2\beta}$ asymptotics, with growth exponents $\beta=1/3$ and $\beta=1/4$, respectively. (c) double log-plot of $W^{2}(\ell,t)$ as a function of $\ell$ for $t \in [5,16]$ with $\Delta t=0.5$ for the Néel (red curves) and Dimer (blue curves) states, compared to the linear slope in $\ell$ (N\'{e}el) and a sub-extensive scaling $\ell^{2\zeta}$ with approximate (fitted) exponent $\zeta \approx 0.38$ (Dimer).}
	\label{fig:iso}
\end{figure*}

\prlsection{Static spin susceptibility}
Dynamical suppression of magnetic fluctuations in SGGEs implies a vanishing static spin susceptibility, namely $\chi=0$, given by the exact formula \cite{DoyonNotes}
\begin{equation}\label{eq:chi}
    \chi = \sum_{s \geq 1} \int \mathrm{d}u \, \chi_{s}(u) [m^{\rm dr}_s(u)]^2,
\end{equation}
where $\chi_{s}(u)\equiv \rho_s(u)(1- n_s(u))$ are the single-mode susceptibilities and $m^{\rm dr}_{s}$ denote the dressed magnetizations of quasiparticles (computed by solving the dressing equations \cite{SM}).
Despite $m^{\rm dr}_{s}$ all vanish upon approaching an unmagnetized state, $q\equiv \lim_{\ell \to \infty}\langle Q_{\ell} \rangle/\ell=0$, absence of uniform convergence requires regularization when evaluating Eq.~\eqref{eq:chi}. In order to reinstate finite magnetization density and finite fluctuations we employ \emph{twisted} initial states. For instance, we use the twisted N\'{e}el state $\ket{\Psi_N (\tau)}=\ket{\psi_{N}(\tau)}^{\otimes L/2}$, where
$\ket{\varphi_{N}(\tau)}\simeq \sum_{\sigma,\sigma'\in \{\uparrow,\downarrow\}}\varphi_{\sigma, \sigma'}\ket{\sigma,\sigma'}$,
with amplitudes $\varphi_{\uparrow \uparrow}(\tau)=-\varphi^{-1}_{\downarrow \downarrow}(\tau)=e^{\tau}$, $\varphi_{\uparrow \downarrow}(\tau)=-\varphi^{-1}_{\downarrow \uparrow}(\tau)=\cot{(\tau/2)}$ depending on `twist' parameter $\tau > 0$. Such twisting in particular ensures that the mode occupation functions experience exponential decay for large $s$, while $m^{\rm dr}_s \sim q\,s^2$ for small/intermediate $s$, mirroring thermal states. Unlike in thermal (Gibbs) states, where the density decay algebraically as $\rho_s(u) \sim s^{-3}$, which gives a finite limit for the susceptibility $\lim_{q \to 0} \chi >0$, SGGEs (in $\Delta>1$ regime) instead generically exhibit exponential falloff $\rho_s(u) \sim e^{-\xi s}$. 
Consequently, the zero-twist limit can be interchanged with the infinite sum over $s$, yielding $\chi=0$.  Such exponential suppression of the densities is not incompatible with the observed freezing of the $n_s$ at large $s$ since the effective Brillouin zone (i.e. the Jacobian $k'_s(u)$) available for quasiparticles with large $s$ shrinks exponentially in SGGEs. Indeed, in this limit only the giant quasiparticles that carry finite effective magnetization, telling that they become effectively pinned locally in space and consequently preclude the distribution of magnetic fluctuations through the system.

\prlsection{Spin diffusion}
We now briefly examine the diffusion constant $\mathfrak{D}$, using the following exact mode resolution \cite{10.21468/SciPostPhys.6.4.049,PhysRevLett.123.186601,PhysRevLett.122.127202} 
 \begin{equation}\label{eq:DiffuConst}
     \mathfrak{D} = \sum_{s \geq 1} \int \mathrm{d}u
     \chi_s(u) |v^{\rm eff}_s(u)| \mu^2_{s},
 \end{equation}
 with effective velocities $v_{s}^{\rm eff}(u)$ (computed from the dressed dispersion relations, see \cite{SM}) and magnetic moments $\mu_{s}\equiv \partial_{q} m_s^{\rm dr}|_{q=0}$. Using the general scaling $\mu_{s} \sim s^2$, alongside $|v^{\rm eff}_s| \sim e^{-\kappa s}$ for $\Delta>1$, we readily conclude that $\mathfrak{D}>0$ in SGGEs. However, as we explain shortly, we find a clear signature of anomalous diffusive behavior. We also note that DC spin conductivity $\sigma$ vanishes identically, that is $\sigma=\mathfrak{D} \chi = 0$.

\prlsection{Numerical simulations}To confirm our theoretical predictions, and to additionally infer the scaling properties of finite subsystems, we carry out numerical simulations using the matrix product states (MPS) with the iTensor library \cite{itensor}. We simulate the time evolution of chains of length $L=100$ up to maximal times $t\approx 15$ (using the maximal bond dimension of $1024$ to achieve convergence in the bond dimension). We compute the time dependence of charge variance in a local subsystem $W^{2}(\ell,t)$ by time-evolving the initial state $\ket{\Psi}$ with TEBD \cite{PhysRevLett.93.040502}, for subsystems of length $\ell$ ranging from $2$ to $40$.
For both N\'{e}el (see Fig.~\ref{fig:sketch}) and Dimer states (see Fig.~\ref{fig:PlotDimerDelta3}), we observe diffusive temporal growth $W^{2}(\ell,t) \sim  t^{1/2}$
followed by saturation to a value $\sim \ell^{2 \zeta}$ with the approximate fitted exponent $\zeta \approx 1/4$ (affording to ignore still very pronounced finite $\ell$ and finite time effects). This value would only be consistent with the FV scaling hypothesis \eqref{eq:growthell} in the case of ballistic dynamical exponent $z=1$, but not with $z=2$ associated with diffusive processes. This leads us to rule out the self-affine structure of the magnetic fluctuations in SGGEs, signalling the absence of normal spin diffusion. Lastly, we also verify that the norm of the equilibrium Hamiltonian $\log \varrho_\ell$ grows super-extensively \cite{SM}, contrasting the extensive behaviour of (quasi)local charges in canonical GGEs (as found e.g. in the XXZ Hamiltonian with $\Delta=0.5$).

\prlsection{Isotropic chain and KPZ fluctuations}
 The isotropic limit $\Delta \to 1$ requires special attention due to an enhanced non-abelian symmetry \cite{Bulchandani_review}. It is by now well-established that in thermal equilibrium with $q=0$ transport of magnetization becomes anomalous, characterized by a super-diffusive dynamical exponent $z=3/2$ characteristic of KPZ physics \cite{PhysRevLett.122.127202,PhysRevLett.122.210602,PhysRevLett.123.186601,Wei2022,PhysRevLett.129.230602,2212.03696,PhysRevE.100.042116}.
 On the other hand, hydrodynamic relaxation from pure states is much less explored, and our work partially fills this void. Indeed, it turns out, somewhat surprisingly, that in the case of isotropic interaction the non-fluctuating initial states can exhibit different qualitative behavior. For example, the N\'{e}el state relaxes to a GGE with regular (i.e. decaying) occupation functions, enabling restoration of fluctuations with a finite $\chi \approx 0.6$. In contrast, the Dimer state again yields $\chi=0$. In fact, using the exact result $n^{\rm D}_{s}(u)=[4u^{2}+(s+1)^{2}]/[(1+4u^{2})(s+1)^{2}]$, the divergence of $\mathfrak{f} = \sum_{s\geq 1}\mathfrak{f}_{s}\to \infty$ follows rigorously from the large-$s$ behavior $\mathfrak{f}_{s}=3\log{(s)}-4\log{(s+1)}+\log{(s+2)}\sim s^{-1}$, implying a logarithmic divergence of $\mathfrak{f}$ with the cutoff $s_{\rm max}$. The limiting  profile for any $u \sim \mathcal{O}(s^{0})$ is the Cauchy--Lorentz distribution, $n^{\rm D}_{\infty}(u)=1/(1+4u^{2})$ \cite{SM}.
 In the case of N\'{e}el state, the spin diffusion constant is found to diverge; the terms in Eq.~\eqref{eq:DiffuConst} tend to constant at large $s$, mirroring the thermal states (exhibiting $\chi_s(u)\sim s^{-3}$ decay, and $|v^{\rm eff}_s(u)| \sim s^{-1}$ (see the additional numerical data in \cite{SM}), stipulated by the `superuniversality' of spin superdiffusion \cite{PhysRevX.11.031023} with dynamical exponent $z=3/2$). As shown in Fig.~\ref{fig:iso}, spin fluctuations grow as $\sim t^{2/3}$, i.e. $\beta=1/3$. In the Dimer quench, however, the equilibrium state reveals distinctly non-thermal features despite preservation of the $SU(2)$ symmetry ($q=0$), with scaling $\chi_s(u) \sim s^{-5}$ and $|v^{\rm eff}_s(u)|\sim s^0$, indicating a logarithmic divergence of $\mathfrak{D}$ with $s_{\rm max}$. Together with the vanishing of the spin susceptibility, this implies finite spin conductivity $\sigma = \chi \mathfrak{D}$, i.e. normal spin transport.

The above findings are well-supported by our numerical simulations. In the N\'{e}el quench, the data is well compatible with the anticipated scaling form with $\zeta=1/2$ and $\beta=1/3$,
\begin{equation}\label{eq:growthNee;}
      W^{2}(\ell,t) \sim \ell \, \varPhi_{\rm iso}\big(t/\ell^{3/2}\big),
\end{equation}
consistently $\chi>0$ and singular spin diffusion constant in the associated GGE. In the Dimer case, we observe (on the accessible times) an algebraic growth $W^{2}(\ell,t)\sim t^{2\beta}$ with $\beta=1/4$, whereas the exponent $\zeta$ appears to be slightly smaller than the extensive value $\zeta=1/2$ (compatibly with the theoretically predicted freezing of $n_{s}$).

\prlsection{Conclusions}By considering a class of non-fluctuating initial product states, we demonstrated that interacting integrable systems hosting infinitely many bound states can evade thermalization to canonical GGEs. We found that local subsystems instead relax towards unorthodox states called squeezed GGEs. The latter feature sub-extensive magnetic fluctuations (signalled by a divergent $U(1)$ chemical potential), and the approach to equilibrium violates the Family-Vicsek scaling hypothesis.  Another distinguished property of SGGEs are non-decaying, so-called `frozen', mode occupations of giant quasiparticles, causing an emergent large-scale semi-classical description \cite{PhysRevLett.125.070601} to break down. We are hopeful that the state-of-the-art quantum simulators \cite{trotzky2012probing,Scherg2021,Wei2022} and modern quantum processors \cite{2209.12889,PhysRevLett.120.050507}
can provide a test bed and an ideal opportunity to verify our predictions, as they do not suffer from the rapid growth of entanglement generated by the quantum quench.

Several recent studies reported anomalous behaviour of macroscopic fluctuating quantities, such as the full counting statistics (FCS) (together with R\'enyi entropy \cite{PhysRevB.99.045150,PhysRevX.12.031016}) of the charge transfer in quantum and classical Heisenberg chains \cite{PhysRevLett.128.090604,gopalakrishnan2022theory,2212.03696} and related models \cite{PhysRevLett.128.160601,krajnik2022universal,PhysRevB.106.205151} featuring fragmentation. It currently remains unclear whether anomalous FCS bears any connection to the observed anomalous dynamic roughening which, according to our simulations, appears to be intimately tied to integrability. Upon breaking integrability, generic Hamiltonians with equidistant energy levels in the limit $\Delta \to \infty$ involve quasi-local quantities conserved up to exponential time $\sim e^{\kappa \Delta}$ for any $\Delta>1$, being a corollary of Ref.~\cite{Abanin2017}. In this view, integrability guarantees the exact conservation of magnetic fluctuations for large subsystems $\ell \gg 1$ even for arbitrary late times. We currently lack any deeper mathematical insight behind this mechanism and how (weak) integrability-breaking perturbations influence the picture, which we plan to investigate in future works.

\prlsection{Acknowledgements}
 We thank L. Zadnik and M. \v{Z}nidari\v{c} for insightful discussions and S. Gopalakrishnan, K. Takeuchi and R. Vasseur for discussions and collaborations on related subjects.
 This work has been partially funded by the ERC Starting Grant 101042293 (HEPIQ) (J.D.N. and G.C.).
 E.I. is supported by project N1-0243 of the Slovenian Research Agency.

\bibliography{apssamp}

\onecolumngrid
\newpage

\appendix
\setcounter{equation}{0}
\setcounter{figure}{0}
\renewcommand{\thetable}{S\arabic{table}}
\renewcommand{\theequation}{S\thesection.\arabic{equation}}
\renewcommand{\thefigure}{S\arabic{figure}}
\setcounter{secnumdepth}{2}

\begin{center}
{\Large Supplementary Material \\ 
\titleinfo
}
\end{center}
\tableofcontents

\section{Additional numerical data}

\begin{figure*}[h!]
	\includegraphics[scale=0.7]{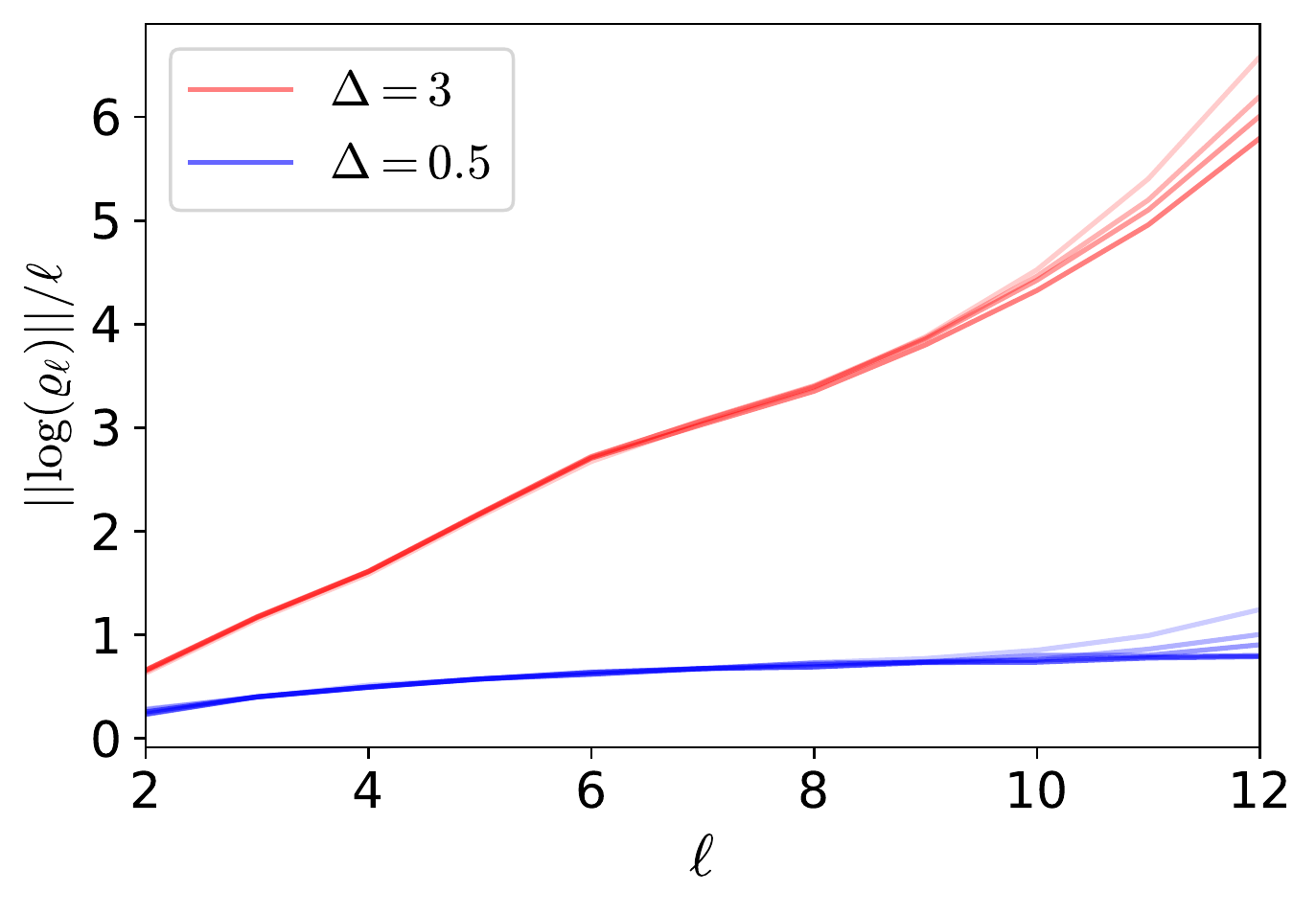}
	\caption{Hilbert--Schmidt norm of the effective equilibrium Hamiltonian $\log \varrho_{\ell}$ of the reduced density matrix $\varrho_\ell(t)$ as function of $\ell \in [2,12]$, shown at different times (increasing from light to dark) in the range $t\in [6,12]$ with step $\delta t=1$. The plot shows that while for $\Delta<1$ the logarithm of the reduced density matrix converges at late time to an extensive operator with whose norm scales as $\sim \ell$, but this is not the case in the regime $\Delta>1$, where the corresponding ensemble is not a canonical but rather an SGGE.}
	\label{fig:log_norm_rho}
\end{figure*}

\begin{figure}[h!]
	\includegraphics[scale=0.6]{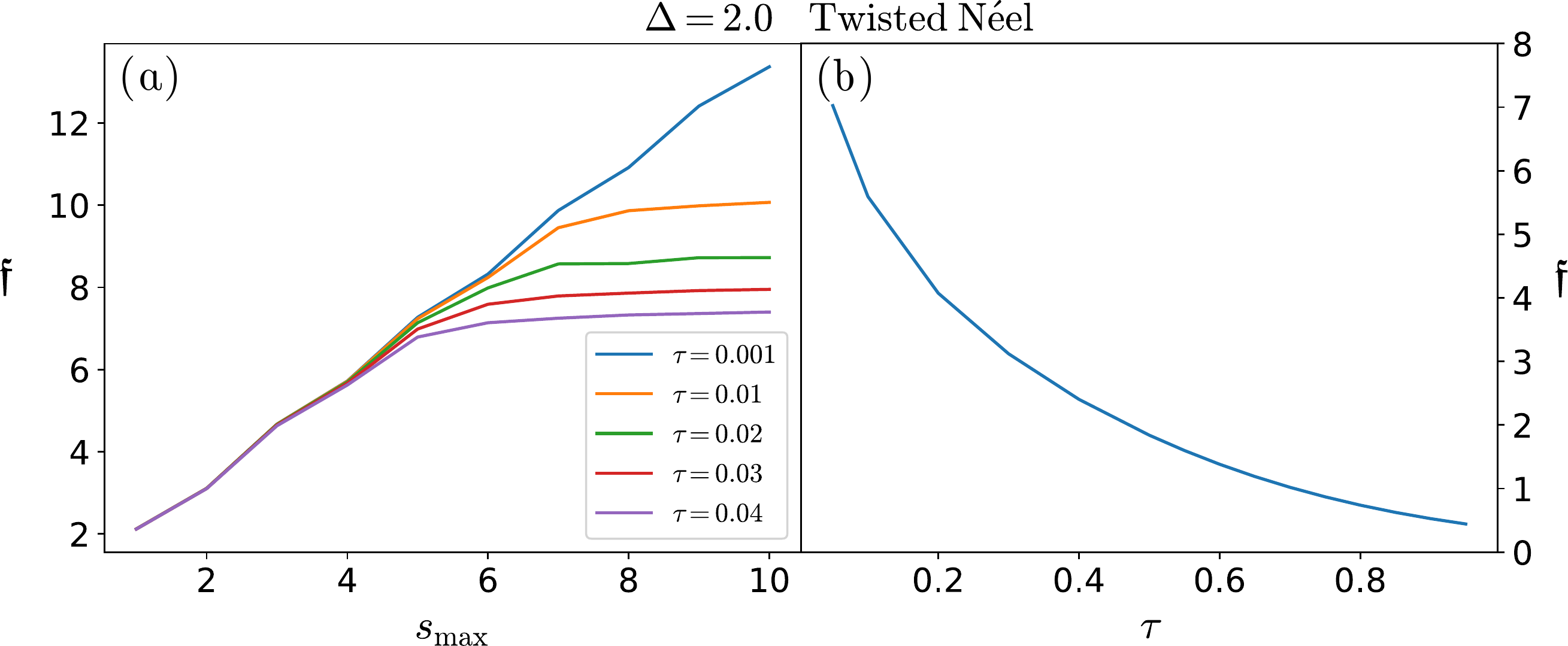}
	\caption{Divergence of $\mathfrak{f}$, representing the quasiparticle contribution  to equilibrium free energy (cf. Eq. 4 in the main text): (a) shown as function of the maximal quantum number of Bethe strings $s_{\rm max}$, for different values of twist parameter $\tau$ in the twisted Néel state and anisotropy $\Delta=2$ (b) as function of deformation $\tau$ by summing over all $s$ (up to numerical precision).}\label{fig:intrho}
\end{figure}

\begin{figure}[h!]
	\includegraphics[scale=0.6]{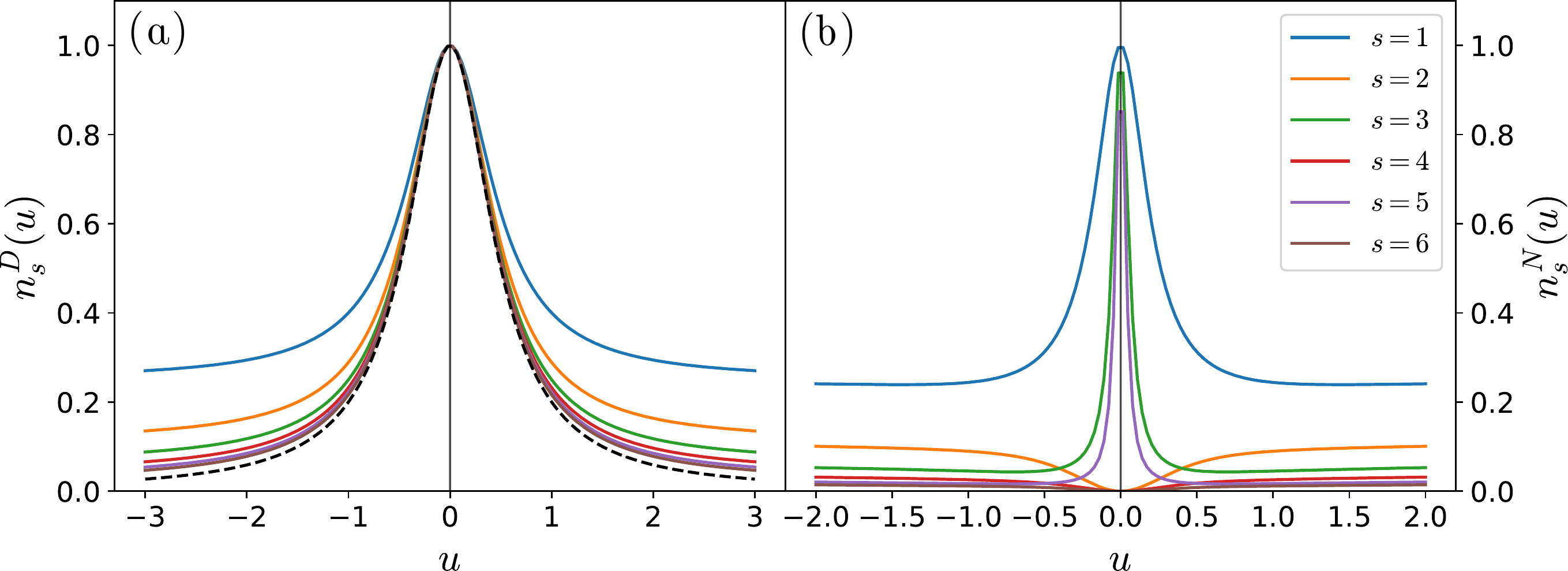}
	\caption{Occupation functions $n_s(u)$ at $\Delta=1$ for a few initial values of $s \in [1,6]$: (left) Dimer state ($u \in [-3,3]$) and (right) N\'{e}el state ($u \in [-2,2]$). Unlike in the N\'{e}el case with generic, i.e. decaying $n_{s}(u)$, the equilibrium ensemble arising from the Dimer quench exhibits freezing, i.e. converge towards the Cauchy--Lorentz distribution $n^{\rm D}_{\infty}(u)=1/(1+4u^{2})$ (black dashed line).
 }\label{fig:intrho}
\end{figure}

\begin{figure}[h!]
	\includegraphics[scale=0.8]{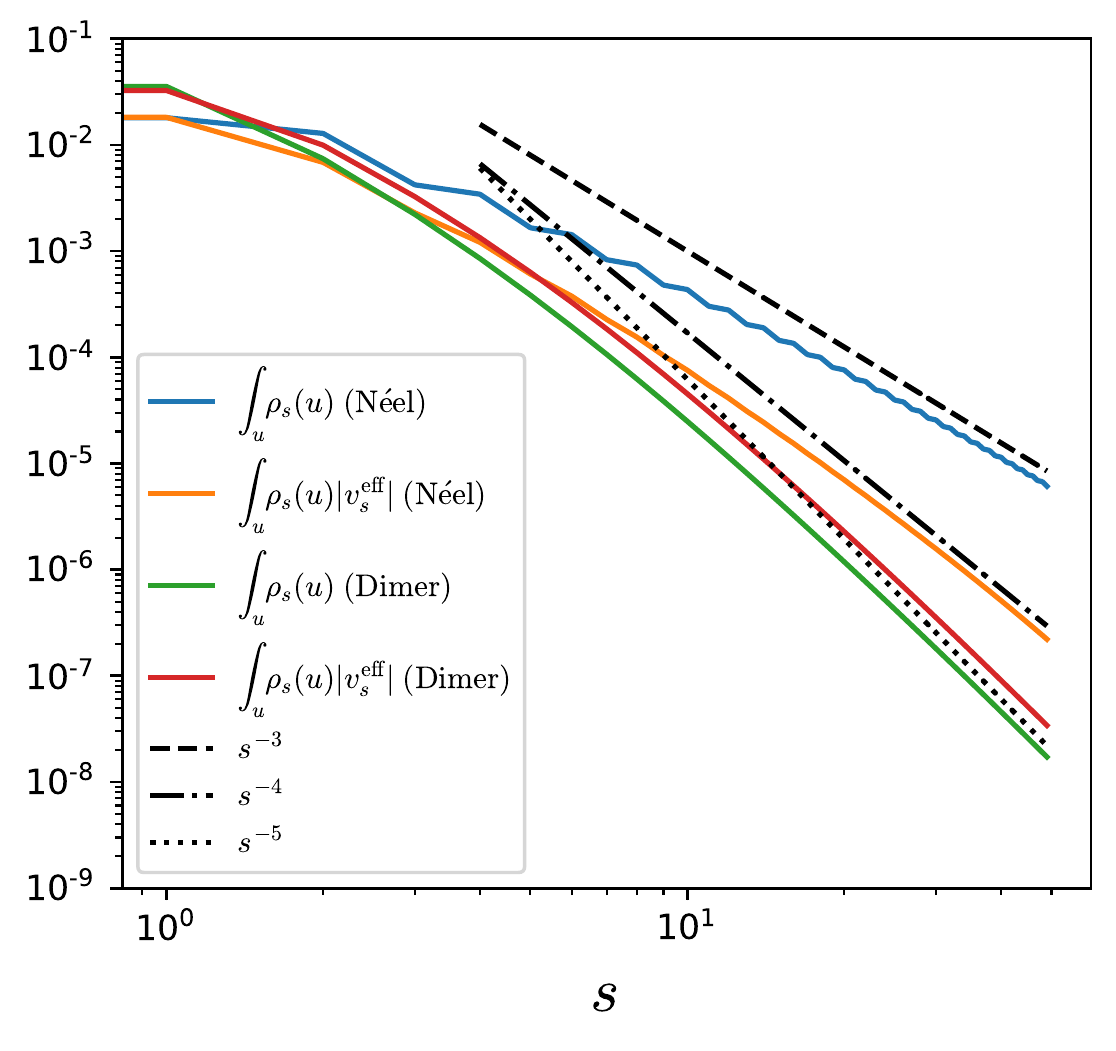}
	\caption{Log-log plot of various rapidity-integrated quantities involving the quasiparticle densities and effective velocities, shown for $\Delta=1$ and  different $s$, including guidelines that indicate different algebraic scaling with $s$. The different exponents are used in the main text to motivate the divergence of the spin diffusion constant or the finiteness or not of the spin susceptibility.
 }\label{fig:intrho}
\end{figure}

\begin{figure}[h!]
	\includegraphics[scale=0.6]{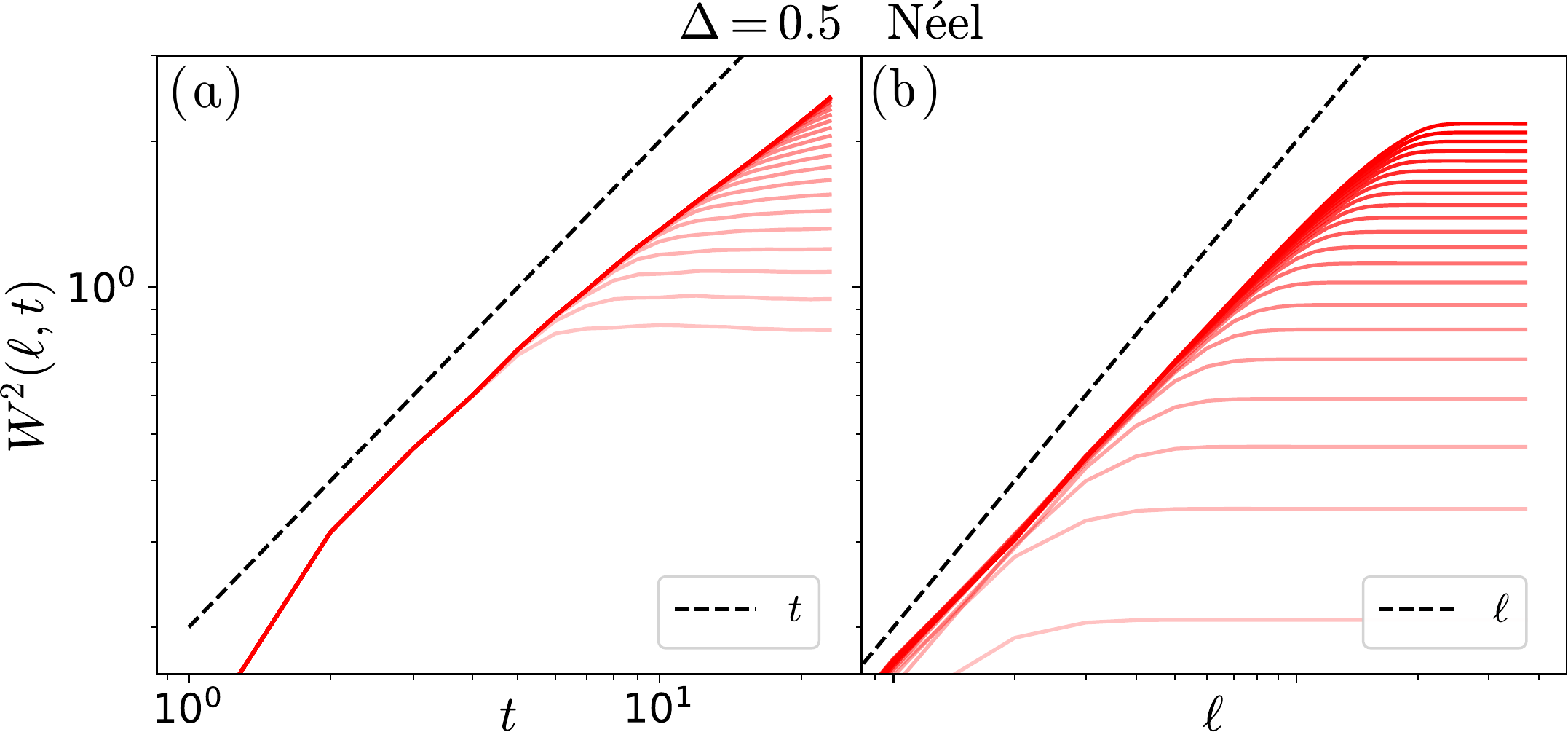}
	\caption{ Temporal and spatial scaling of magnetic fluctuations after a quench from the Néel state in the Heisenberg XXZ chain with anisotropy $\Delta=0.5$ (where ballistic scaling is expected): (a) double log-plot of $W^{2}(\ell,t)$ as a function of time for different $\ell \in [4,40]$ (ranging from light to dark); (b) double log-plot of $W^{2}(\ell,t)$ as a function of $\ell$ for different times $t \in [5,15]$ with $\delta t=0.5$ (from light to dark), compared to the expected linear asymptotic scaling $W^{2}(\ell,t) \sim \ell$.
 }\label{fig:intrho}
\end{figure}

\begin{figure}[h!]
	\includegraphics[scale=0.6]{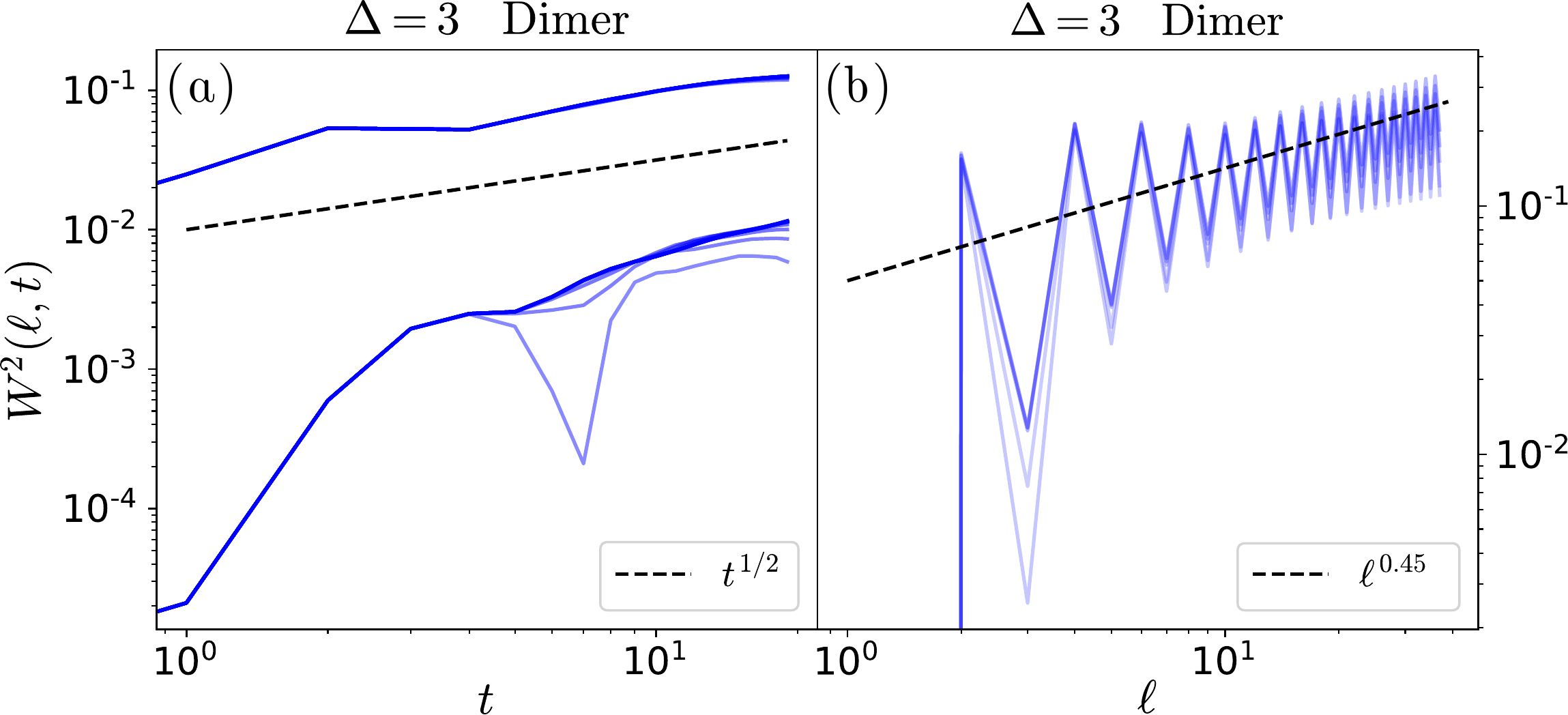}
	\caption{ Temporal and spatial scaling of magnetic fluctuations after a quench from the Dimer state in the Heisenberg XXZ chain with anisotropy $\Delta=3$ : (a) double log-plot of $W^{2}(\ell,t)$ as a function of time for different $\ell \in [4,40]$ (ranging from light to dark); (b) double log-plot of $W^{2}(\ell,t)$ as a function of $\ell$ for different times $t \in [12,15]$ with $\delta t=0.5$ (from light to dark), with the fitted asymptotic scaling $W^{2}(\ell,t) \sim \ell^{0.45}$ (closely analogous to the Néel initial state case in the main text).
 }\label{fig:PlotDimerDelta3}
\end{figure}

\begin{figure}[h!]
	\includegraphics[scale=0.6]{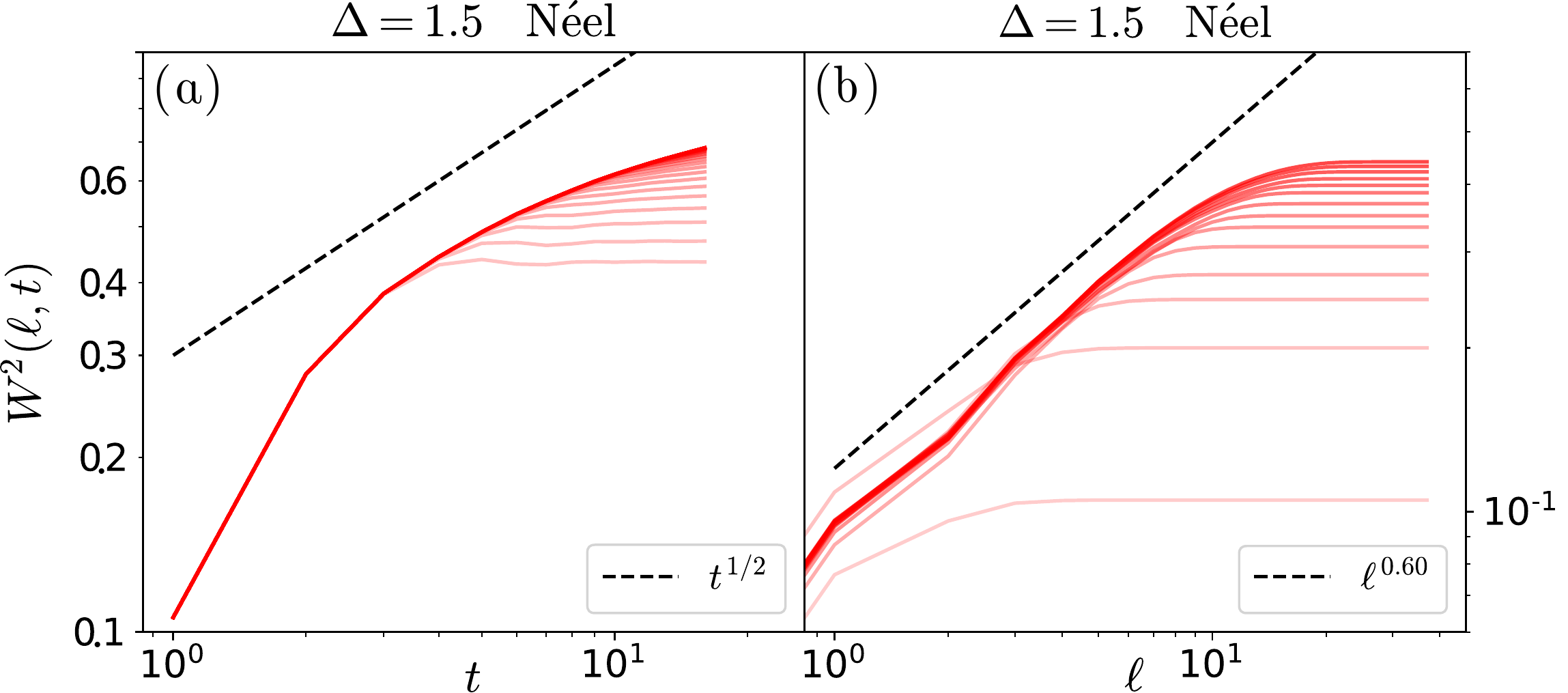}
	\caption{ Temporal and spatial scaling of magnetic fluctuations after a quench from the Dimer state in the Heisenberg XXZ chain with anisotropy $\Delta=1.5$ : (a) double log-plot of $W^{2}(\ell,t)$ as a function of time for different $\ell \in [4,40]$ (ranging from light to dark); (b) double log-plot of $W^{2}(\ell,t)$ as a function of $\ell$ for different times $t \in [12,15]$ with $\delta t=0.5$ (from light to dark), with the fitted asymptotic scaling $W^{2}(\ell,t) \sim \ell^{0.45}$ (closely analogous to the Néel initial state case in the main text).
 }\label{fig:PlotDimerDelta3}
\end{figure}

\clearpage

\section{Generalized Gibbs ensembles in the Heisenberg chain}

Here we provide a compressed summary of generalized Gibbs ensembles in the anisotropic Heisenberg XXZ spin chain in the language of Thermodynamic Bethe Ansatz. Specializing to regime $|\Delta|\geq 1$, we discuss the analytic properties of various state functions and derive the formulae for the free-energy density.
For definiteness, we consider the isotropic interaction $\Delta = 1$ first, and later in a separate section discuss appropriate modifications to describe the model with $\Delta>1$.
We shall make extensive use the following compact notations for the scalar product and convolution,
\begin{equation}
g \circ h \equiv \sum_{s}\int_{\mathbb{R}} \mathrm{d} u g_{s}(u)h_{s}(u),\qquad
g \star h \equiv \sum_{s}\int_{\mathbb{R}} \mathrm{d} u g_{s}(u-u')h_{s'}(u'),
\end{equation}
respectively, for any set of real (dummy) functions $f,g,h$ supported on the real rapidity axis $u\in \mathbb{R}$. Moreover, we employ a compact notation for imaginary shifts by $\pm \ii/2$, that is $g^{\pm}(u)\equiv g(u\pm \tfrac{\ii}{2})$ and $g^{[\pm n]}(u)\equiv g(u\pm n\tfrac{\ii}{2})$.

\medskip

\paragraph{Commuting fused transfer matrices.}
Considering the unitary irreducible spin-$s/2$ representations $\mathcal{V}_{s}$ of dimension $s+1$
with spin generators $\hat{S}^{\alpha}_{s}$,
the quasilocal conserved charges of the spin-$1/2$ Heisenberg chain on $L$ sites with the Hilbert space $\mathcal{H}\cong \mathcal{V}^{\otimes L}_{1}$, are derived from the fused transfer matrices $\hat{T}_{s}(u)$,
\begin{equation}\label{eq:T}
    \hat{T}_{s}(u) = {\rm Tr}_{\mathcal{V}_{s}}\bigotimes_{L}\hat{L}_{s,1}(u),
\end{equation}
where
\begin{equation}
    \hat{L}_{s,1}(u) = u\,\mathds{1} + 2\ii \sum_{\alpha \in \{x,y,z\}}\hat{S}^{\alpha}_{s}\otimes \hat{S}^{\alpha}_{1},
\end{equation}
are quantum Lax operators acting on the tensor product $\mathcal{V}_{s}\otimes \mathcal{V}_{1}$,
while traces in Eq.~\eqref{eq:T} are over the common auxiliary space $\mathcal{V}_{s}$, $u \in \mathbb{C}$ is a general complex spectral parameter.
Mutual commutativity of $\hat{T}_{s}(u)$
 \begin{equation}
  [\hat{T}_{s}(u),\hat{T}_{s'}(u')]=0,\qquad s,s' \in \mathbb{Z}_{\geq 0},
 \end{equation}
 with $\hat{T}_{0}(u)=u^{L}$, is ensured by the Yang--Baxter equations. For more details, we direct the reader to \cite{PhysRevLett.115.120601,ilievski2016quasilocal}. 

\medskip

\paragraph{Quasilocal charges.}
The quasilocal conserved charges correspond to logarithmic derivatives, that is
\begin{equation}
\hat{X}_{s}(u) = \frac{1}{2\pi \ii}\partial_{u}
\log{\frac{\hat{T}^{+}_{s}(u)}{\phi_{s}(u)}},
\end{equation}
where $\phi_{s}(\theta)$ is a convenient normalization,
and $u \in \mathcal{P}$ lies in the physical strip
\begin{equation}
\mathcal{P} = \{u \in \mathbb{C};|{\rm Im}(u)|<1/2\}.
\end{equation}

In the thermodynamic limit, magnon rapidities $\{u_{i}\}$ (Bethe roots) organize into Bethe strings,
\begin{equation}
\bigcup_{i=1}^{M}\{u_{i}\} \rightarrow
\bigcup_{j=1}^{\infty}\bigcup_{i=1}^{M_{j}}\bigcup_{k=1}^{j}\{u_{j,i}+\tfrac{\ii}{2}(j+1-2k)\}, 
\end{equation}
representing bound-state excitations. Generic thermodynamic eigenstates involve $M=\sum_{s}M_{s}$ magnons, partitioned into $M_{s}\sim \mathcal{O}(L)$ strings (up to corrections that are exponentially small in $L$).

Bound state undergo elastic collisions. The elementary scattering amplitude of two magnons is simple rational function depending on the rapidity difference $u$,
\begin{equation}
    S(u) = \frac{u-\ii}{u+\ii},
\end{equation}
The fused scattering amplitude associated with a $j$-string and a $k$-string read
\begin{equation}
    S_{j,\ell}(u) = \prod_{a=1}^{j}\prod_{b=1}^{\ell}S(u+(j-\ell-2a+2b)\tfrac{\ii}{2}),
\end{equation}
with convention that $S(0)=-1$. The differential scattering phases defined the set of kernels
\begin{equation}
    K_{j,\ell}(u) = \frac{1}{2\pi \ii}\partial_{u}\log S_{j,k}(u).
\end{equation}

In the large-$L$ limit, the separation between nearby strings with rapidities $u_{i}\sim \mathcal{O}(L^{0})$ decreases as $1/L$, allowing to introduce densities of Bethe strings $\rho_{s}(u)$.
The Bethe equations
\begin{equation}
e^{\ii k(u_{i})L}\prod_{j=1}^{M}S(u_{i}-u_{j}) = -1,    
\end{equation}
can then be formulated as the Bethe--Yang integral equations
\begin{equation}
    \rho_{s} + \bar{\rho}_{s} = K_{s} - K_{s,s'}\star \rho_{s'},
\end{equation}
where
\begin{equation}
   K_{s}(u) = \frac{1}{2\pi}k^{\prime}_{s}.
\end{equation}

\medskip

\paragraph{String-charge duality.}
Expressed in terms of the root densities, eigenvalues of $\hat{X}_{s}(u)$ read
\begin{equation}\label{eq:Xrho}
X_{s} = G_{ss'}\star \rho_{s'},
\end{equation}
where
\begin{equation}
    G_{s,s'} = \sum_{j=1}^{{\rm min}(s,s')}K_{s+s'+1-2j},
\end{equation}
represents the `bare energy tensor'. In terms of the $s$-kernel,
\begin{equation}
s(u) = \frac{1}{2\cosh{(\pi u)}}.
\end{equation}
we can write it as
\begin{equation}
   G_{s,s'} = (1+K)_{s,s'}\star s. 
\end{equation}
Eq.~\eqref{eq:Xrho} is referred to as the string-charge duality. By introducing the discrete d'Alembertian operator $\square$, acting on any set of dummy functions $g_{s}$ as
\begin{equation}
(\square g)_{s} = \square_{ss'}g_{s'} =
(s^{-1}\delta_{s,s'} - I_{s,s'})\star g_{s'},
\end{equation}
the inverse of Eq.~\eqref{eq:Xrho} reads
\begin{equation}
\rho_{s} = \square X_{s}.    
\end{equation}
We have simultaneously introduced the infinite dimensional incidence matrix $I_{s,s'}$ of ${\rm A}_{\infty}$ root system,
\begin{equation}
I_{s,s'} \equiv \delta_{s,s'-1} + \delta_{s,s'+1}.
\end{equation}
Notice that the left inverse $s^{-1}$ of the convolution kernel $s(u)$, $s^{-1}\star s = \delta$, acts as an imaginary shift in the complex $\theta$-plane, which has to be taken with the $\epsilon$-prescription,
\begin{equation}
(s^{-1}\star g)(\theta) = g(u+\tfrac{\ii}{2}-\ii \epsilon) + g(u-\tfrac{\ii}{2}+\ii \epsilon),
\end{equation}
in order to avoid touching the boundary of the physical strip $\mathcal{P}$.
In fact, $s^{-1}$ is the pseudoinverse since it involves a non-trivial nullspace, i.e. there exist functions $\zeta$ such that $s^{-1}\star \zeta=0$.

\medskip

\paragraph*{Analytic parametrization.}
Following Ref.~\cite{ilievski2017from}, we being by writing the reduced density matrix of a GGE for a spin chain of length $L$ in the form
\begin{equation}
\hat{\varrho} = \mathcal{Z}^{-1}_{L}\exp{\left[-\sum_{s=1}^{\infty}\int \dd u \lambda_{s}(\theta)\hat{X}_{s}(\theta)-h\,\hat{S}^{z}_{\rm tot}\right]},
\end{equation}
where functions $\lambda_{s}(\theta)$ play the role of Lagrange multipliers. The distinguished global $U(1)$ charge,i.e. total magnetization, couples to chemical potential $h$.

As pointed out in Ref.~\cite{ilievski2017from}, the above representation of a GGE is not completely general. Indeed, the corresponding thermodynamic $\mathcal{Y}$-functions (to be introduced shortly) will always be \emph{holomorphic} in the physical strip $\mathcal{P}$ in the complex $\theta$-plane. Generic $\mathcal{Y}$-functions are however meromorphic, i.e. they involve isolated zeros and poles in the interior of $\mathcal{P}$. To accommodate for this, one has to include the charges $\hat{X}_{s}(u)$ with complex arguments $u\in \mathcal{P}$, combined into conjugate pairs to ensure hermiticity. Accordingly, by changing the basis and introducing conserved operators
\begin{equation}
\hat{\rho}_{s}=\square \hat{X}_{s},
\end{equation}
the GGE density matrix can be cast in the form 
\begin{equation}
\hat{\rho}_{\rm GGE} = \mathcal{Z}^{-1}_{L} \exp{\left[-\mu_{s}\circ \hat{\rho}_{s}-h\,\hat{S}^{z}_{\rm tot}\right]}.
\end{equation}
Now $\mu_{s}(\theta)$ can be interpreted as fugacities ascribed to quasiparticles, i.e. magnonic bound states. In terms of original Lagrange multipliers, we have the relation
\begin{equation}
\mu_{s} = G_{ss'}\star \lambda_{s'} + h\,s.
\end{equation}

\medskip

\paragraph{Thermodynamic Bethe Ansatz.}
In the limit of large system size $L$, the finite-volume partition sum
$\mathcal{Z}_{L} = {\rm Tr}\,\hat{\varrho}_{\rm GGE}$
exhibits exponential growth,
\begin{equation}
\mathcal{Z}_{L} \asymp e^{-L\,f},
\end{equation}
where
\begin{equation}
f = -\lim_{L\to \infty}\frac{1}{L}\log \mathcal{Z}_{L},
\end{equation}
is the free energy per site. To compute $f$, we employ the saddle-point technique called the Thermodynamic Bethe Ansatz. To this end, we begin by casting the partition function as a functional integral,
\begin{equation}
\mathcal{Z}[\{\rho_{s}\}] \equiv \int \mathcal{D}[\{\rho_{s}\}]e^{-L\,\mathcal{F}[\{\rho_{s}\}]},
\end{equation}
in terms of the free-energy functional
\begin{equation}
\mathcal{F}[\{\rho_{s}\}] = \mathcal{E}[\{\rho_{s}\}] - \mathcal{S}_{\rm YY}[\{\rho_{s}\}]
+ h\left[\frac{1}{2}-\sum_{s}\int \dd u\,s\,\rho_{s}(u)\right],
\end{equation}
with
\begin{equation}
\mathcal{E}[\{\rho_{s}\}] = \sum_{s}\int \dd u\,\mu_{s}(u)\rho_{s}(u),\qquad
\mathcal{S}_{\rm YY}[\{\rho_{s}\}] = \sum_{s}\int \dd u \, \mathfrak{s}_{s}(u),
\end{equation}
where each mode contributes to the density of entropy a weight
\begin{equation}
\mathfrak{s}_{s}(u) = (\rho_{s}(u)+\bar{\rho}_{s}(u))\log{\big[(\rho_{s}(u)+\bar{\rho}_{s}(u))\big]}
-\rho_{s}(u)\log{(\rho_{s}(u))} - \bar{\rho}_{s}(u)\log{(\bar{\rho}_{s}(u))}.
\end{equation}
The variational variables are only the root densities $\rho_{s}$ represent, whereas the hole densities $\bar{\rho}_{s}$ are related to $\rho_{s}$
via the Bethe--Yang integral equations,
\begin{equation}
\rho_{s} + \bar{\rho}_{s} = K_{s} - K_{ss'}\star \rho_{s'},
\end{equation}
and thus we have
\begin{equation}
\delta \bar{\rho}_{s} = -(1+K)_{ss'}\star \delta \rho_{s'}.
\end{equation}
Free-energy density is obtained by variational optimization of the free-energy functional $\mathfrak{F}$, that is
\begin{equation}
f=\mathcal{F}[\{\rho_{s}\}]_{\rho_{s}=\rho^{*}_{s}},
\end{equation}
where $\rho^{*}_{s}$ represent the saddle-point densities.
Affording a slight abuse of notation, we shall subsequently operate only with the saddle-point quantities and omit the star symbol.
By introducing the thermodynamic $\mathcal{Y}$-functions,
\begin{equation}
    \mathcal{Y}_{s}(u) \equiv \frac{\bar{\rho}_{s}(u)}{\rho_{s}(u)},
\end{equation}
and rewriting the entropic contribution compactly as
\begin{equation}
\mathfrak{s}_{s} = \rho_{s}\log{[1+\mathcal{Y}_{s}]} + \bar{\rho}_{s}\log{[1+1/\mathcal{Y}_{s}]},
\end{equation}
the saddle-point condition
\begin{equation}
 \delta \mathcal{F}/\delta \rho_{s}|_{\rho^{*}_{s}}=0,   
\end{equation}
yields, upon substituting $\delta \bar{\rho}_{s}$ with $\delta \rho_{s}$,
\begin{equation}
0 = \mu_{s} - s\,h + \log{(1+\mathcal{Y}_{s})} - (1+K)_{s,s'}\star \log{(1+1/\mathcal{Y}_{s'})}.
\end{equation}
These saddle-point equations are the celebrated \emph{canonical TBA equations}
\begin{equation}
\log \mathcal{Y}_{s} = \mu_{s} - h\,s + K_{ss'}\star \log(1+1/\mathcal{Y}_{s'}).
\end{equation}

The logarithm of $\mathcal{Y}$-functions determined the dressed quasiparticle energies via
\begin{equation}
    \varepsilon_{s}(u) = \log \mathcal{Y}_{s}(u).
\end{equation}
The effective velocity of quasiparticle propagation
are the computed based on the dressed dispersion relations, namely
\begin{equation}
v_{s}^{\rm eff}(u) = \frac{\partial \varepsilon_{s}}{\partial p_{s}} = \frac{\varepsilon_{s}^{\prime}(u)}{p_{s}^{\prime}(u)},
\end{equation}
where dressed momenta $p^{\prime}_{s}(u)$ coincide with the total state densities,
\begin{equation}
    p^{\prime}_{s}(u) = 2\pi(\rho_{s}+\bar{\rho}_{s}).
\end{equation}

\medskip

\paragraph{Free energy.}
In terms of the thermodynamic $\mathcal{Y}$-functions, the density of free energy can be expressed as
\begin{equation}
f = \frac{h}{2} + (\mu_{s}-h\,s)\circ \rho_{s}-\rho_{s}\circ \log{(1+\mathcal{Y}_{s})}
- \bar{\rho}_{s}\circ \log{(1+1/\mathcal{Y}_{s})}.
\end{equation}
Using the Bethe--Yang equations $\bar{\rho}_{s}=K_{s}-(1+K)_{ss'}\star \rho_{s'}$, we can eliminate $\bar{\rho}_{s}$ and recast the above expression in the form
\begin{equation}
f = \frac{h}{2} + (\mu_{s}-h\,s - \log{(1+\mathcal{Y}_{s})})\circ \rho_{s}
- K_{s}\circ \log{(1+1/\mathcal{Y}_{s})} +(1+K)_{ss'}\star \rho_{s'}\circ\log{(1+1/\mathcal{Y}_{s})}.
\end{equation}
The last term can be simplified using the symmetry $K_{ss'}(\theta,\theta')=K_{s's}(\theta',\theta)$, yielding
\begin{equation}
(1+K)_{ss'}\star \rho_{s'}\circ \log{(1+1/\mathcal{Y}_{s})} = \rho_{s}\circ (1+K)_{ss'}\star \log(1+1/\mathcal{Y}_{s'})
= \rho_{s}\circ \log{(1+\mathcal{Y}_{s})} - \rho_{s}\circ (\mu_{s} - h\,s),
\end{equation}
where in the second line we have made use of the TBA equations $K_{ss'}\star \log{(1+1/\mathcal{Y}_{s'})}=\log \mathcal{Y}_{s}-(\mu_{s} - hs)$.
This means that in the above expression for $f$ two terms get cancelled out, leaving us with the simple result
\begin{equation}
f = \frac{h}{2} -K_{s}\circ \log{(1+1/\mathcal{Y}_{s})}.
\end{equation}
The outlined derivation only makes use of kernel identities is thus completely \emph{general} and free of any analyticity assumptions.

\medskip

\paragraph{Resumation.}
We now show how to resolve the infinite summation to obtain a simple compact expression for $f$.
To this end, we insert the resolution of the identity $(1-R)_{ss''}\star(1+K)_{s''s}=\delta_{ss'}$ and rewrite $f$ as
\begin{equation}
f = \frac{h}{2} -K_{s}\circ (1-R)_{ss'}\star \big[(1+K)_{ss'}\star \log{(1+1/\mathcal{Y}_{s'})}\big].
\end{equation}
The term in the square bracket can now again be simplified by using the TBA equations,
$\log{(1+\mathcal{Y}_{s})}=\mu_{s} - h\,s+(1+K)_{ss'}\star \log{(1+1/\mathcal{Y}_{s'})}$, yielding
\begin{equation}
f = \frac{h}{2} -K_{s}\circ (1-R)_{ss'}\star \big[\log{(1+\mathcal{Y}_{s'})}-\mu_{s'} + h s'\big].
\end{equation}
Using further that $(1-R)_{ss'}\star \alpha\,s'=0$ for any constant $\alpha$, we arrive at
\begin{equation}
f^{*} = \frac{h}{2} -\big[\log{(1+\mathcal{Y}_{s})}-\mu_{s}\big]\circ (1-R)_{ss'}\star K_{s'}.
\end{equation}
Here we encounter the following subtlety: the convolution kernel $(1-R)$ is not the proper inverse of Fredholm kernel $(1+K)$ since it possesses a non-trivial null space.
Evidently, upon convolving with $(1-R)$ the $h$-dependent term in the square bracket gets \emph{erased}. In effect, the resulting expression for $f^{*}$ does in fact not equal $f$.
To better elucidate this point, notice that $\lim_{s\to \infty}\mu_{s}=0$, along with $\log \mathcal{Y}_{s}(h)\sim -h\,s$ at large $s$, imply that $\lim_{s\to \infty}[\log{(1+\mathcal{Y}_{s'})}-\mu_{s'} + h s'\big]=0$. On the other hand, 
after acting with the convolution we ended up with $\lim_{s\to \infty}s^{-1}[\log{(1+\mathcal{Y}_{s})}-\mu_{s}\big]=-h$.
This issue can be elegantly circumvented by introducing regularized $\mathcal{Y}$-functions,
\begin{equation}
\mathcal{Y}^{\rm reg}_{s} = \frac{\mathcal{Y}_{s}}{\mathcal{Y}^{\infty}_{s}},\qquad \mathcal{Y}^{\infty}_{s} = e^{-h\,s},
\end{equation}
with unit asymptotics, $\lim_{s\to \infty}\mathcal{Y}^{\rm reg}_{s}=1$ writing
\begin{equation}
\log(1+\mathcal{Y}_{s}) - \mu_{s} + h\,s = \log{(1+1/\mathcal{Y}_{s})} + \log \mathcal{Y}^{\rm reg}_{s} - \mu_{s},
\end{equation} 
such that we can now safely convolve with $(1-R)$ without losing any information, yielding
\begin{equation}
f = \frac{h}{2} -[\log(1+1/\mathcal{Y}_{s})+\log \mathcal{Y}^{\rm reg}_{s} - \mu_{s}] \circ (1-R)_{ss'}\star K_{s'}s.
\end{equation}
By finally using the identity $(1-R)_{ss'}\star K_{s'}=\delta_{s,1}s$, we obtain
\begin{equation}
f = \frac{h}{2} + s\circ \big(\mu_{1}
-\log{(1+1/\mathcal{Y}_{1})}\big) + s\circ \log \mathcal{Y}_{1} - h\circ s,
\end{equation}
which, upon cancelling the first term with the last one using $1 \circ s = 1/2$, brings us to the final compact result
\begin{equation}\label{eq:f_compact}
f = s\circ \big(\mu_{1}-\log{(1+\mathcal{Y}_{1})}\big).
\end{equation}
Remarkably, there is no explicit $h$-dependence in this formula, unlike in the original (canonical) formula.
Let us also stress again that we have performed all the computations independently of any state-specific information. Formula \eqref{eq:f_compact} follows purely from the formal structure of the canonical TBA equations and kernel identities stemming from the underlying fusion rules.

\medskip

For practical purposes, it is beneficial to replace the initial source term $\mu_{1}(\theta)$ with the local source term $d_{1}(u)$.
This can be achieved via the identity
\begin{equation}
s\circ \mu_{1} = d_{1}\circ [(1+K_{2})\star s],
\end{equation}
which follows as a consequence of $\mu_{1}=(1+K_{1,1})\star d_{1}$. Using further that $(1+K_{2})\star s = K_{1}$, we arrive at
\begin{equation}
f = K_{1}\circ d_{1} - s\circ \log{(1+\mathcal{Y}_{1})}.
\end{equation}
The key advantage of this formula is that now $d_{1}$ can be retrieved entirely
from the analytic data of the initial $\mathcal{Y}$-function without ever needing to compute $\mu_{1}$.

\medskip

\paragraph*{Decoupled TBA equations.}

In order to fully exhibit the analytic structure of the $\mathcal{Y}$-functions, we proceed by recasting the TBA equations in the so-called decoupled form.
To this end, we rewrite the canonical TBA equations in the form
\begin{equation}
(1+K_{ss'})\star \log \mathcal{Y}_{s'} = \mu_{s} - h\,s + K_{ss'}\star \log{(1+\mathcal{Y}_{s'})},
\end{equation}
and by subsequently applying the pseudo-inverse $(1-R)$ we obtain the equivalent decoupled form
\begin{equation}
\log \mathcal{Y}_{s} = d_{s} + I_{ss'}s\star \log{(1+\mathcal{Y}_{s'})},
\end{equation}
with local source terms
\begin{equation}
d_{s} = (1-R)_{ss'}\star \mu_{s'}.
\end{equation}
Here we again encounter the ``nullspace problem''; we have to ensure the solution to the decoupled equations matches the original solution of the canonical equations. To achieve this, we request the large-$s$ dependence
\begin{equation}
\mathcal{Y}_{s} \sim e^{-h\,}s.
\end{equation}
The local source terms admit the following general decomposition,
\begin{equation}
d_{s} = s\star \lambda_{s} + \zeta_{s},
\end{equation}
where $\zeta$-functions lie entirely in the nullspace of the pseudo-inverse $s^{-1}$, namely
\begin{equation}
s^{-1}\star \zeta_{s} = 0.
\end{equation}

Finally, we recast the local TBA equations into equivalent functional relations, written in terms of complex $\mathcal{Y}$-functions obtained via
analytically continuation from the real rapidity axis. The procedure goes as follows. Operating first on both sides by $s^{-1}$ and taking the exponent, we readily obtain
\begin{equation}
\mathcal{Y}^{+}_{s}(u-\ii \epsilon)\mathcal{Y}^{-}_{s}(u-\ii \epsilon)
= e^{\lambda_{s}(u)}\log{[(1+\mathcal{Y}_{s-1}(u))(1+\mathcal{Y}_{s+1}(u))]},
\end{equation}
using the usual compact notation for imaginary shift $g^{\pm}(u)\equiv g(u \pm \tfrac{\ii}{2})$.
An infinitesimal regulator $\epsilon$ is required here to avoid touching the boundary of the physical strip $\mathcal{P}$. The above equations, which generally involve state-dependent node terms, and called the modified $\mathcal{Y}$-system.

The local TBA equations can be transformed back to the canonical TBA equations as follows. First, the $\mathcal{Y}$-functions appearing
on left-hand side of
\begin{equation}
\log \mathcal{Y}_{s} - I_{ss'}s\star \log \mathcal{Y}_{s'} = d_{s} + I_{ss'}s\star \log{(1+1/\mathcal{Y}_{s'})},
\end{equation}
are replaced with the regularized ones and written as $(1-R)_{ss'}\star \log \mathcal{Y}^{\rm reg}_{s'}$. Next, convolving both sides
with $(1+K)$ and using $(1+K)_{s,s''}\star I_{s'',s'}s = K_{ss'}$ yields
\begin{equation}
\log \mathcal{Y}^{\rm reg}_{s} = \mu_{s} + K_{ss'}\star \log{(1+1/\mathcal{Y}_{s'})},
\end{equation}
where
\begin{equation}
\mu_{s} = (1+K)_{ss'}\star d_{s'}.
\end{equation}
Lastly, one replaces $\mathcal{Y}^{\rm reg}_{s}$ with $e^{h\,s}\mathcal{Y}_{s}$ to recover the missing term $-h\,s$.

\subsection*{Anisotropic easy-axis regime}

The construction from the previous section  extends with little modifications. Parametrizing anistotropy as $\Delta=\cosh{(\eta)}$ with $\eta \in \mathbb{R}$,
the Brillouin zone now wraps to a circle of circumference $\pi$. We pick the fundamental zone, i.e. work with rapidities in the range $u \in (-\pi/2,\pi/2)$. Analytic continuation
to complex $u$ now extends to $|{\rm Im}(u)|<\eta/2$, i.e. the physical strip is a cylinder
$\mathcal{P}_{\eta}\equiv (-\pi/2,\pi/2)\times (-\ii \eta/2,\ii \eta/2)$. Compactification of rapidities implies the $k$-space becomes a discrete lattice, with the forward and backward discrete Fourier transforms reading
\begin{equation}
f(u) = \frac{1}{\pi}\sum_{k\in \mathbb{Z}}\hat{f}(k)e^{-\ii\,2k u},\qquad
\hat{f}(k) = \int^{\pi/2}_{-\pi/2}\dd u f(u) e^{\ii\,2u k}.
\end{equation}

The fundamental kernel inversion identity is structurally preserved under $\eta$-deformation,
\begin{equation}
(1+K)^{-1} = (1-R),
\end{equation}
except that this time the Fredholm resolvent $R = I_{s,s'}\mathfrak{s}$ involves a more complicated $\mathfrak{s}$-kernel, namely a doubly-periodic function. Its Fourier representation is however very simple,
\begin{equation}
\hat{\mathfrak{s}}(k) = \frac{1}{2\cosh{(k\eta)}},
\end{equation}
and therefore we can immediately give a series representation 
\begin{equation}
\mathfrak{s}(u) = \frac{1}{\pi}\sum_{k\in \mathbb{Z}}\frac{e^{\ii\,2k u}}{2\cosh{(k\eta)}}.
\end{equation}

To derive the $\mathcal{Y}$-system functional relations, we require analytic continuation to the boundaries of $\mathcal{P}_{\eta}$. In analogy with the isotropic case, we introduce the discrete shift, denoted by $s^{-1}$, once again taken with the $\epsilon$-prescription
\begin{equation}
[s^{-1}_{\eta}\star f](u) = \lim_{\epsilon \to 0}\big[f(u + \tfrac{\ii \eta}{2}-\ii \epsilon) + f(u - \tfrac{\ii \eta}{2} + \ii \epsilon)\big],
\end{equation}
Analogously to the undeformed case, we have
$s^{-1}\star \mathfrak{s}=\delta$ and
thus the following useful identity,
\begin{equation}
s^{-1}\star (\mathfrak{s}\star f) = f \qquad {\rm for}\quad u \in (-\pi/2,\pi/2),
\end{equation}
which holds for any dummy function $f$. This follows immediately from $\pm \ii \eta/2$ shifts
and by recalling the resolution of the Dirac delta $\delta(u)=\tfrac{1}{\pi}\sum_{k\in \mathbb{Z}}e^{\ii\,2k u}$.

To find the rapidity space representation of kernel $\hat{\mathfrak{s}}(k)$, we make use of the Jacobi elliptic functions.
To this end, we introduce the elliptic integral
\begin{equation}
K(m) = \int^{\pi/2}_{0}\frac{\dd \varphi}{\sqrt{1-m\sin^{2}(\varphi)}},\qquad K^{\prime}(m) \equiv K(1-m),
\end{equation}
and recall the series expansion
\begin{equation}
\frac{2K}{\pi}{\rm dn}(\upsilon) = 1 + 4\sum_{n\geq 1}\frac{\cos{(2n \tfrac{\pi}{2K}\upsilon)}}{q^{n}+q^{-n}}, 
\end{equation}
with nome $q=\exp{(-\pi K^{\prime}/K)}$. Setting $q=e^{\eta}$ and introducing $u = (\pi/2K)\upsilon$,
we can rewrite it as a Fourier series,
\begin{equation}
\mathfrak{s}(u) = \frac{1}{2\pi} \frac{2K}{\pi} {\rm dn}\Big(\frac{2K}{\pi}u\Big)
= \frac{1}{2\pi}\left[1+2\sum_{k\geq 1}\frac{\cos{(2ku)}}{\cosh{(k\eta)}}\right]
= \frac{1}{\pi}\sum_{k\in \mathbb{Z}}\frac{\cos{(2ku)}}{2\cosh{(k\eta)}}.
\end{equation}

For generic macrostates, the $\mathcal{Y}$-system relations are modified by the presence of additional node data, namely
\begin{equation}
\mathcal{Y}^{+}_{s}\mathcal{Y}^{-}_{s} = e^{\lambda_{s}}(1+\mathcal{Y}_{s-1})(1+\mathcal{Y}_{s+1}).
\end{equation}
Taking the logarithm and subsequently undoing the contour shifts by convolving with $\mathfrak{s}$, we arrive at a local form of the TBA equations
\begin{equation}
\log \mathcal{Y}_{s} = d_{s} + \log{[(1+\mathcal{Y}_{s-1})(1+\mathcal{Y}_{s+1})]},
\end{equation}
with source terms $d_{s}$ related to $\lambda_{s}$ via
\begin{equation}
d_{s} = \mathfrak{s} \star \lambda_{s} + \zeta_{s}.
\end{equation}
In the same way as in the isotropic case, the local source terms $d_{s}$ pick up additional nullspace contributions $\zeta_{s}$,
\begin{equation}
s^{-1}_{\eta}\star \zeta_{s} = 0,
\end{equation}
whenever the $\mathcal{Y}$-function contains a zero or a pole inside $\mathcal{P}$. On the other hand, Lagrange multipliers that determine the so-called node data instead span the kernel of $s^{-1}_{\eta}$ and are therefore retrieved by virtue $s^{-1}_{\eta}\star (\mathfrak{s}\star \lambda_{s})=\lambda_{s}$.
We must accordingly find which doubly-periodic functions span ${\rm ker}(s^{-1}_{\eta})$. In addition, in order to be able to factor the analytic data out of $\mathcal{Y}_{s}$, we need the $\eta$-deformed $\tau$-function, say $\tau_{\eta}(u;w)$ to obey the identity $\tau^{+}(u-w)\tau^{-}(u-w)=1$, such that in the $\eta\to 0$ limit we recover
$\tau_{\eta}(u;w)\to \tau(u;w) = \tanh{(\tfrac{\pi}{2}(u-w))}$. Moreover, by sending $w$ towards the strip boundary
we have to retrieve $\mathfrak{s}$. 

\medskip

To deduce the explicit form of the basis functions for nullspace of $s^{-1}_{\eta})$, it will suffice to have a brief look at the thermal Gibbs case.
The associated partition function corresponds to the large-$N$ limit of $\mathcal{Z}^{(N)}_{L}$ associated with a classical vertex model in the torus topology with physical circumference $L$ and perpendicular (fictitious) dimension $N$, understood as the Trotter parameter. The general construction of the isotropic Heisenberg model can be found in Ref.~\cite{10.21468/SciPostPhys.7.3.033} which, mostly for compactness of presentation, we do not review here.

The analytic structure of the initial $\mathcal{Y}$-function, denoted by $\mathcal{Y}^{(N)}_{1}$, is inherited from the leading eigenvalue of the fundamental column transfer matrix. For finite $N$,
the initial $\mathcal{Y}$-function (describing the truncated Gibbs state) involves a zero of degree $2N$ located at $\theta=\pm\ii(\eta/2 \mp \alpha_{N})$ and no other analytic data inside $\mathcal{P}$, with $\alpha_{N}\equiv \beta/2N$.
In the large-$N$ limit, the solution converges to the Gibbs state with the local source term coupling to the initial node becomes
\begin{equation}
 d^{\rm Gibbs}_{1}(\theta)=-4\pi\beta\,\mathfrak{s}(\theta).   
\end{equation}
This can be easily seen by performing the limit in Fourier space,
\begin{align}
-2\pi N \frac{1}{\pi}\sum_{k\in \mathbb{Z}}e^{-2\ii k \theta}\frac{\sinh{(2k \alpha_{N})}}{k\cosh{(k\eta)}}
&=-4\pi \beta \frac{1}{\pi}\sum_{k\in \mathbb{Z}}\frac{\sinh{(2k\alpha_{N})}}{2k \alpha_{N}}\hat{\mathfrak{s}}(k) \to -4\pi \beta \mathfrak{s}(k),
\end{align}
showing that finite-$N$ regularization is in fact the discrete approximation to the Dirac $\delta$-function. We thus conclude that the initial $\lambda$-function of the GGE,
namely $\lambda_{1}(u) \simeq \delta(u)$. Recalling that $\mu_{1} \simeq\lambda_{1}\star \mathfrak{s}$, we infer that
$s^{-1}_{\eta}\star \mathfrak{s}=\delta$ fall outside the nullspace of $s^{-1}_{\eta}$.
This explains the general mechanism of how the node data arises from the analytic data of $\mathcal{Y}$-functions $\mathcal{Y}^{(N)}_{s}$ in the scaling limit. In this regard, it is necessary that
a degree $N$ zero or pole approaching in a $1/N$ fashion towards the boundary $\partial \mathcal{P}_{\eta}$; contributions from isolated zeros of poles of finite order would otherwise be negligible.
The upshot of the above analysis is that any individual zero or a pole located at $w\in \mathcal{P}_{\eta}$ (restricted to the upper half, ${\rm Im}(w)>0$) contributes a term
\begin{equation}
\hat{\zeta}(k;v) = -2\pi \frac{\sinh(-2\ii k v)}{k} = 2\pi\ii \frac{\sin{(2k v)}}{k},
\end{equation}
where the parameter $v \in \mathcal{P}_{\eta}$ is related to $w$ via
\begin{equation}
w = \frac{\ii \eta}{2} - v,\qquad {\rm Im}(v) \leq \frac{\eta}{2}.
\end{equation}
We note that $v$ represents an inhomogeneity of the column transfer matrix. The thermal Gibbs case is a special case where zeros are located at $w_{N} = \ii \eta/2 - \ii \alpha_{N}$, corresponding to $v_{N}=\ii \alpha_{N}$.

To give a concrete example, we briefly consider the Dimer state. The corresponding $\mathcal{Y}$-functions involve a double zero at the origin ($v=\ii \eta/2$) and a double zero at the edge of the Brillouin zone at $u=\pm \eta/2$ ($v=\ii \eta/2 + \pi/2$) for all the nodes $s \in \mathbb{N}$. This results in the source term of the form
\begin{equation}
d^{\rm D}_{s}(u) = \frac{1}{\pi}\sum_{k\in \mathbb{Z}}\hat{d}^{\rm D}_{s}(k),
\end{equation}
with
\begin{equation}
\hat{d}^{\rm D}_{s}(k) = [\hat{\zeta}(k;\ii \eta/2) + \hat{\zeta}(k;\ii \eta/2+\pi/2)]\hat{\mathfrak{s}}(k)
=\sum_{k\in \mathbb{Z}}\frac{\tanh{(k\eta)}}{k}\Big(-1-(-1)^{k}\Big).
\end{equation}

\section{Thermodynamic $\mathcal{Y}$-system}

Here we summarize how to compute the thermodynamic $\mathcal{Y}$-functions $\mathcal{Y}_{s}(u)$ from the expectation values of the quasilocal charges $\hat{X}_{s}(u)$ on the initial state $\ket{\Psi}$.

We employ the transfer matrix method, assuming the state has a product form, i.e. can be written as $\ket{\Psi}=\ket{\psi}^{\otimes L/b}$ with period $b \in \mathbb{N}$. First, we define a family of the double-row transfer matrices
\begin{equation}
    \mathbb{T}^{\Psi}_{s}(u;\alpha) \equiv \bra{\psi}\bigotimes_{b}\mathbb{L}_{s}(u;\alpha)\ket{\psi},
\end{equation}
with Lax operators
\begin{equation}
    \mathbb{L}_{s}(u;\alpha) \equiv \frac{L^{-}_{s}(u)L^{+}(u+\alpha)}{L^{[-s-1]}_{0}(u)L^{[s+1]}_{0}(u+\alpha)},
\end{equation}
acting of the tensor product $\mathbb{C}^{2}\otimes \mathcal{V}^{\otimes 2}_{s}$, where $\mathcal{V}_{s}$ is an auxiliary unitary spin-$s/2$ representation.
We have introduced an additional parameter $\alpha \in \mathbb{C}$ formally acting as source coupling to $\hat{X}_{s}(u)$.
The partition function $\mathcal{Z}^{\Psi}_{s}(u;\alpha)$ can be computed by iteration, yielding
\begin{equation}
    \mathcal{Z}^{\Psi}_{s}(u;\alpha) = \lim_{L\to \infty}\frac{1}{L}{\rm Tr}_{\mathcal{V}^{\otimes 2}_{s}}[\mathbb{T}^{\Psi}_{s}(u;\alpha)]^{L/b}.
\end{equation}
The average values of quasilocal charges
\begin{equation}
    X^{\Psi}_{s}(u) = \lim_{L\to \infty}\frac{1}{L}\bra{\Psi}\hat{X}_{s}(u)\ket{\Psi},
\end{equation}
are therefore given by
\begin{equation}
    X^{\Psi}_{s}(u) = -\ii \partial_{\alpha}\mathcal{Z}^{\Psi}_{s}(u;\alpha)|_{\alpha=0}.
\end{equation}

The information stored in the set of complex functions $\{X^{\Psi}_{s}(u)\}$ with $u \in \mathcal{P}$ uniquely fixes an equilibrium macrostate. By employing the string-charge duality \cite{ilievski2016string}, the root and hole densities of Bethe strings are given by
\begin{align}
    \rho^{\Psi}_{s}(u) &= \square X^{\Psi}_{s}(u) = X^{\Psi}_{s}(u+\tfrac{\ii}{2}) + X^{\Psi}_{s}(u-\tfrac{\ii}{2}) - X^{\Psi}_{s-1}(u) - X^{\Psi}_{s+1}(u),\\
    \bar{\rho}^{\Psi}_{s}(u) &= K_{s}(u) - X^{\Psi}_{s}(u+\tfrac{\ii}{2}) - X^{\Psi}_{s}(u-\tfrac{\ii}{2}),
\end{align}
from where we readily obtain
\begin{equation}
    \mathcal{Y}^{\Psi}_{s}(u) = \frac{\bar{\rho}^{\Psi}_{s}(u)}{\rho^{\Psi}_{s}(u)}.
\end{equation}

For meromorphic solutions, i.e. macrostate whose $\mathcal{Y}$-functions only involve the analytic data (zeros and poles inside the physical strip $\mathcal{P}$) and no node data (cf. Refs.~\cite{ilievski2017from,10.21468/SciPostPhys.7.3.033}), the thermodynamic $\mathcal{Y}$-function obey the canonical $\mathcal{Y}$-system relations (without additional node terms). In this case, it suffices to infer the initial $\mathcal{Y}$-function, while the higher ones can be obtained in a closed analytic form simply by iteration, namely
\begin{equation}\label{eq:Ysystem}
    \mathcal{Y}_{s}(u) = \frac{\mathcal{Y}^{+}_{s-1}(u)\mathcal{Y}^{-}_{s-1}(u)}{1+\mathcal{Y}_{s-2}(u)} - 1,\qquad s \geq 2.
\end{equation}
In particular, the entire family of two-site integrable boundary states, representing generic product states with $b=2$, always yields such meromorphic solutions.
In general, states with $b=2$ involve also non-trivial node data. In such cases, an iterative solution of the $\mathcal{Y}$-system hierarchy is no longer possible and one has to compute the $\mathcal{Y}$-functions one by one.

\subsection*{Numerical solution of the $\mathcal{Y}$-system}
The recursive relation \eqref{eq:Ysystem} allows obtaining all $\mathcal{Y}$-functions once we know explicitly the first one for $s=1$.  Focusing first on the gapped regime, $\Delta = \cosh{(\eta)}$ with $\eta >0$, we employ the outlined procedure to deduce $\mathcal{Y}_{1}(u)$, with $u \in [-\pi/2,\pi/2]$
(suppressing additional dependence on $\eta$ and twisting parameters).
\begin{itemize}
\item Dimer state $\ket{\rm D} =\Big[\frac{\ket{\uparrow \downarrow} - \ket{\downarrow \uparrow}}{\sqrt{2}}\Big]^{\otimes L/2}$, with
\begin{equation}
\mathcal{Y}^{\rm D}_{s}(u)=\frac{1}{2} \tan ^2(u) \csc (u-\ii \eta ) \csc (u+\ii \eta ) (\cosh (2 \eta )+3 \cos (2 u)+2).
\end{equation}
\item N\'eel state $\ket{\rm N} = \ket{\uparrow \downarrow}^{\otimes L/2}$, with
\begin{align}
\mathcal{Y}^{\rm N}_{s}(u) &=
\frac{1}{8} \sin ^2(2 u) \csc (u-\ii \eta ) \csc (u+\ii \eta ) \sec (u-\ii \eta ) \sec (u+\ii \eta ) \nonumber \\
&\times \text{csch}\left(\frac{1}{2} (\eta -2 \ii u)\right)
   \text{csch}\left(\frac{1}{2} (\eta +2 \ii u)\right) (\cosh (\eta )+2 \cosh (3 \eta )-3 \cos (2 u)).
\end{align}
\item One-parameter family of twisted N\'{e}el states
\begin{equation}
\ket{{\rm N}(\tau)}=
\frac{1}{\mathcal{N}_{{\rm N}(\tau)}}
\left[
e^{\tau}\ket{\uparrow \uparrow} + \cot{(\tau/2)}\ket{\uparrow \downarrow} - \tan{(\tau/2)}\ket{\downarrow \uparrow} - e^{-\tau}\ket{\downarrow \downarrow}\right],    
\end{equation}
with normalization $\mathcal{N}_{{\rm N}(\tau)}=2\cosh{(2\tau)}+\cot^{2}{(\tau/2)}+\tan^{2}{(\tau/2)}$,
recovering $\ket{\rm N}$ as $\tau\to 0$.
The first $\mathcal{Y}$-function is quite lengthy and we thus do not display it here.
\item Two-parameter family of twisted Dimer states
\begin{equation}
\ket{{\rm D}(\tau;\gamma)}=
\frac{1}{\mathcal{N}_{{\rm D}(\tau;\gamma)}}
\left[
\frac{e^{\gamma}}{\cosh{(\eta)}}\ket{\uparrow \uparrow} + \frac{\ii}{\sinh{(\eta/2)}}\ket{\uparrow \downarrow} - \frac{\ii}{\sinh{(\eta/2)}}\ket{\downarrow \uparrow} - \frac{e^{-\gamma}}{\cosh{(\eta)}}\ket{\downarrow \downarrow}\right],    
\end{equation}
with normalization $\mathcal{N}_{{\rm D}(\tau;\gamma)}=\sqrt{2}\sqrt{({\rm csch}^{2}(\eta/2)+\cosh{(2\gamma)}{\rm sech}^{2}{(1/\tau)})}$,
recovering the Dimer state $\ket{\rm D}$ as $\tau \to 0$. The initial $\mathcal{Y}$-function reads
\begin{align*} 
32 e^{4\gamma}(1+\mathcal{Y}^{{\rm D}(\tau;\gamma)}_{s}(u)) &=
-\sec ^2(u) \csc (u-\ii \eta ) \csc (u+\ii \eta ) \sec ^2\left(u-\frac{\ii}{\tau }\right) \sec ^2\left(u+\frac{\ii}{\tau }\right) (\cos (4 u)-\cosh (2 \eta )) \nonumber \\
&\times
   \left(\left(1+e^{4\gamma}\right) \cosh (\eta )+2 e^{2\gamma} \cosh \left(\frac{2}{\tau }\right)-e^{4\gamma} \cos (2 u-\ii \eta )-\cos (2 u-\ii \eta )+2 e^{2\gamma}\right) \nonumber \\
   &\times \left(\left(1+e^{4\gamma}\right)
   \cosh (\eta )+2 e^{2\gamma} \cosh \left(\frac{2}{\tau }\right)-e^{4\gamma} \cos (2 u+\ii \eta )-\cos (2 u+\ii \eta )+2 e^{2\gamma}\right).
\end{align*}
In our applications, we set $\gamma=1$.

\end{itemize}
To numerically generate the higher $\mathcal{Y}$-functions we use the following procedure:
\begin{enumerate}
    \item We create a matrix $Y_{i,1,k} = \mathcal{Y}_{1}(u_i + (N_{\rm shift}/2 - k) \ii \eta/2 ) $ where the rapidities $u$ are discretized by Gaussian quadrature with $N_{\rm point}$, the integer $k \in [0,N_{\rm shift}-1]$, and with the integer $s \in [0,S_{\rm{max}}]$.
    \item Using the convention that $\mathcal{Y}_0(u)=0$, we then calculate the corresponding values starting on the plane $s=1$ and $s=2$, and generate for $s \geq 2$ as 
\begin{equation}
    Y_{i,j,k} =  \frac{Y_{i,j-1,k-1} Y_{i,j+1,k+1}}{1+Y_{i,j-2,k}} - 1.
\end{equation}
    \item In the matrix $Y_{i,j,k}$, we select the plane corresponding to $k=N_{\rm shift}/2$.
\end{enumerate}

We remind there is a one-to-one correspondence between the occupation (filling) functions and the $\mathcal{Y}$-functions:
\begin{equation}
	n_s(u)= \frac{1}{1+\mathcal{Y}_s(u)}.
\end{equation}

\end{document}